\documentclass[pra,aps,twocolumn,superscriptaddress,nofootinbib,nopacs]{revtex4} 

\usepackage{amsmath}
\usepackage{amssymb}
\usepackage{amsfonts}
\usepackage{bm}
\usepackage{braket}
\usepackage{color}
\usepackage{dcolumn}
\usepackage{epsfig}
\usepackage{gensymb}
\usepackage{graphicx}
\usepackage{lmodern}
\usepackage{mathrsfs}
\usepackage{mathtools}
\usepackage{pst-all}
\usepackage{psfrag}
\usepackage{units}
\usepackage[colorlinks,linkcolor=blue,citecolor=blue,urlcolor=blue,hyperindex,driverfallback=dvipdfm]{hyperref}

\def\ii{{\rm i}}  \def\ee{{\rm e}}
  
\def\rb{{\bf r}}  \def\Rb{{\bf R}}    
\def\xx{\hat{\bf x}}  \def\yy{\hat{\bf y}}  \def\zz{\hat{\bf z}}

\def\kb{{\bf k}}  \def\kpar{k_\parallel}  \def\kparb{{\bf k}_\parallel}
    
  \def\kB{{k_{\rm B}}}
\def\Eb{{\bf E}}      
  
\def\rp{r_{\rm p}}  
\def\vF{v_{\rm F}}  \def\kF{{k_{\rm F}}}  \def\EF{{E_{\rm F}}}
      \def\kp{k_{\rm p}}  
  
\newcommand{\thetav}{\boldsymbol{\theta}}
\newcommand{\uv}{\textbf{u}}
\newcommand{\Eepsv}{\vec{\mathcal{E}}}
\def\vF{v_{\rm F}}  \def\kF{{k_{\rm F}}}  \def\EF{{E_{\rm F}}}  \def\w{\omega}  \def\qv{{\bf q}}
 \def\ce{{c_{\rm e}}}  \def\nF{{n_{\rm F}}}
\def\Te{{T_{\rm e}}}   \def\Tl{{T_l}}
  \def\cl{{c_{l}}}
\def\Rp{\mathcal{R}_{\rm p}}

\newcommand{\PreserveBackslash}[1]{\let\temp=\\#1\let\\=\temp}
\newcolumntype{C}[1]{>{\PreserveBackslash\centering}p{#1}}
\newcolumntype{R}[1]{>{\PreserveBackslash\raggedleft}p{#1}}
\newcolumntype{L}[1]{>{\PreserveBackslash\raggedright}p{#1}}
\def\cw{1.3cm}
\def\cww{1.7cm}

\newcommand{\abs}[1]{\mathopen{}\left|#1\right|\mathclose{}}

\def\II{{\mathbb{I}}}

\newcommand{\eps}{\epsilon}

\newcommand{\vv}{\textbf{v}}

\newcommand{\Ev}{\mathbf{E}}
\newcommand{\Eeps}{\mathcal{E}}

\newcommand{\kvp}{\mathbf{k}_\parallel}

\newcommand{\Rv}{\mathbf{R}}

\begin{document} 
\title{Ultrafast Momentum-Resolved Probing of Plasmon Thermal Dynamics with Free Electrons}

\author{Vahagn~Mkhitaryan}
\thanks{These two authors contributed equally to the work.}
\affiliation{ICFO-Institut de Ciencies Fotoniques, The Barcelona Institute of Science and Technology, 08860 Castelldefels (Barcelona), Spain}

\author{Eduardo~J.~C.~Dias}
\thanks{These two authors contributed equally to the work.}
\affiliation{ICFO-Institut de Ciencies Fotoniques, The Barcelona Institute of Science and Technology, 08860 Castelldefels (Barcelona), Spain}

\author{Fabrizio~Carbone}
\affiliation{Laboratory for Ultrafast Microscopy and Electron Scattering (LUMES), Institute of Physics, \'Ecole Polytechnique F\'ed\'erale de Lausanne (EPFL), Lausanne CH-1015, Switzerland}

\author{F.~Javier~Garc\'{\i}a~de~Abajo}
\email{javier.garciadeabajo@nanophotonics.es}
\affiliation{ICFO-Institut de Ciencies Fotoniques, The Barcelona Institute of Science and Technology, 08860 Castelldefels (Barcelona), Spain}
\affiliation{ICREA-Instituci\'o Catalana de Recerca i Estudis Avan\c{c}ats, Passeig Llu\'{\i}s Companys 23, 08010 Barcelona, Spain}

\begin{abstract}
Current advances in ultrafast electron microscopy make it possible to combine optical pumping of a nanostructure and electron beam probing with sub{\aa}ngstrom and femtosecond spatiotemporal resolution. We present a theory predicting that this technique can reveal a rich out-of-equilibrium dynamics of plasmon excitations in graphene and graphite samples. In a disruptive departure from the traditional probing of nanoscale excitations based on the identification of spectral features in the transmitted electrons, we show that measurement of angle-resolved, energy-integrated inelastic electron scattering can trace the temporal evolution of plasmons in these structures and provide momentum-resolved mode identification, thus avoiding the need for highly-monochromatic electron beams and the use of electron spectrometers. This previously unexplored approach to study the ultrafast dynamics of optical excitations can be of interest to understand and manipulate polaritons in 2D semiconductors and other materials exhibiting a strong thermo-optical response.
\end{abstract}
\maketitle
\date{\today}

\section{Introduction}

Thermal engineering of plasmons and other forms of polaritons in nanomaterials offers an appealing way of controlling light-matter interactions down to nanometer \cite{BQ12} and femtosecond \cite{NWG16,paper345} spatiotemporal scales, opening applications in photonics and optoelectronics, such as all-optical switching \cite{DUS18,paper337}, light modulation \cite{LCM14,AAF15,paper313}, ultrafast light emission \cite{KGS18_2}, and photodetection \cite{paper315}. Traditionally, the study of ultrafast thermal dynamics relies on optical experiments, in which a light pump pulse is used to excite the system and bring it out of equilibrium, followed by a light probe pulse that measures the evolution of the sample response \cite{LH01,GSD08,WWF12}. However, this procedure is limited in spatial resolution due to light diffraction when relying on far-field optics, or to a few tens of nanometers when a tip is used to locally amplify the electromagnetic field in ultrafast scanning near-field optical microscopy (SNOM) \cite{NWG16}.

Electron energy-loss spectroscopy (EELS) performed in scanning transmission electron microscopes overcomes the optical diffraction limit by using $30-300\,$keV electrons rather than light to map the material response \cite{paper149,KS14,paper338} with sub{\aa}ngstrom spatial precision \cite{BDK02} and increasing spectral resolution that currently enables the study of mid-infrared polaritons \cite{KLD14,LTH17,HNY18,HKR19,HHP19}. When the electron beam is well collimated, momentum-resolved inelastic electron scattering grants us access into the dispersion relations of surface modes in planar films \cite{BGI1966,PSV1975,CS1975,CS1975_2}, while the dispersion of thicker samples can be probed with lower spatial resolution through low-energy ($\sim50-500\,$eV) electron microscopy in reflection mode \cite{R95,NHH01}. Additionally, a combination of high temporal and spatial resolution has been achieved through the development of ultrafast electron microscopy, based on the use of femtosecond light and electron pulses that are simultaneously aimed at the sample with a well-controlled relative delay \cite{GLW06,BPK08,BFZ09}. By scanning the light frequency, this approach additionally brings meV energy resolution in what is known as electron energy-gain spectroscopy (EEGS) \cite{H99,paper114,H09}, which has been experimentally demonstrated \cite{paper306} to challenge the state-of-the-art benchmark of a few meV achieved through tour-de-force advances in electron optics \cite{KLD14}. In the photon-induced near-field electron microscopy (PINEM) technique \cite{BFZ09,paper151,PLZ10,PLQ15,FES15,paper282,EFS16,PRY17,paper311,paper332,KLS20,WDS20}, the electron beam is focused with nanoscale spatial precision, while the relative light-electron delay provides femtosecond temporal resolution. PINEM has been used to shoot femtosecond movies from surface plasmons evolving in nanowires \cite{PLQ15} and buried interfaces \cite{paper282}, and more recently, also in the characterization of optical dielectric cavities \cite{KLS20,WDS20}. Although efforts in this context have emphasized light-matter interaction aspects and our ability to modulate the wave function of free-space electrons, the optical-pump/electron-probe (OPEP) approach has strong potential to study nanoscale dynamics with unrivalled spatiotemporal resolution by addressing material properties that range from relatively slow structural \cite{GLW06,BPK08} and electronic \cite{VSH18} behavior to the intrinsically ultrafast nonlinear optical response \cite{paper347}.

Two-dimensional (2D) materials offer a splendid testbed for OPEP because they generally undergo substantial changes in their electronic structure under optical pumping. We consider in particular highly-doped graphene, which in addition hosts electrically-tunable plasmons \cite{JBS09,FAB11,paper176,paper196,FRA12,YLC12,YLL12,BJS13} that possess long lifetime \cite{WLG15,NMS18}, strong spatial confinement \cite{LGA17,AND18}, and a large nonlinear response \cite{KKG13,YLH18,KDM18,paper337}. These properties have prompted the exploration of exciting applications that include infrared photodetection \cite{XML09,KMA14,LGW17,paper315,paper346}, optical sensing \cite{paper256,HYZ16,paper319}, and light modulation \cite{paper235,GSG15,YLH18,paper337,paper345}. Because of its conical electronic band structure, the thermo-optical response is remarkably high in graphene and manifests in the emergence of plasmons in heated undoped samples \cite{V06_2,paper235,NWG16}, as well as plasmon shifts when the electronic temperature is increased \cite{JKW16,paper313}. The effects are dramatic at electronic temperatures of a few 1000s\,K, which can be reached using femtosecond laser pulses without damaging the material \cite{JUC13,GPM13}. In this context, while SNOM has been extensively used to characterize graphene plasmons \cite{FAB11,paper196,FRA12,NWG16,NMS18}, the unique combination of space, time, momentum, and energy resolution offered by OPEP makes it an ideal technique to reveal unexplored properties of those excitations, as well as other types of polaritons and their associated electron/lattice dynamics in 2D materials.

Here, we use predictive theory to demonstrate that the ultrafast OPEP approach can be used to characterize the temporal dynamics of plasmons in both extended and nanostructured graphene and graphite films. Specifically, we show that the strong confinement of plasmons in these materials produces large deflection in the inelastically scattered electrons, directly yielding dispersion curves in the energy-momentum-resolved electron transmission maps. Adjustment of the light/electron delay allows us to explore the temporal evolution of these excitations as the material undergoes an initial rapid increase in electronic temperature upon optical pumping, followed by slower cooling through relaxation to the atomic lattice over a subpicosecond timescale. Importantly, for laterally confined plasmons, such as transverse modes in ribbons, there is a strong correlation between plasmon energy and momentum, which enables the identification of these modes by collecting the angle-resolved transmitted electrons integrated over a wide energy window, thus avoiding the need to use a spectrometer. This approach is particularly advantageous to study low-energy modes, where conventional imaging in the Fourier plane of an electron microscope could serve to identify polaritons in a spectral window below the accessible range in currently available setups. The present results should stimulate the use of OPEP to study the ultrafast dynamics of polaritions in materials that possess a strong thermo-optical response, such as graphene and other 2D crystals in extended and nanostructured geometries.

\begin{figure*}
\centering
\includegraphics[width=\textwidth]{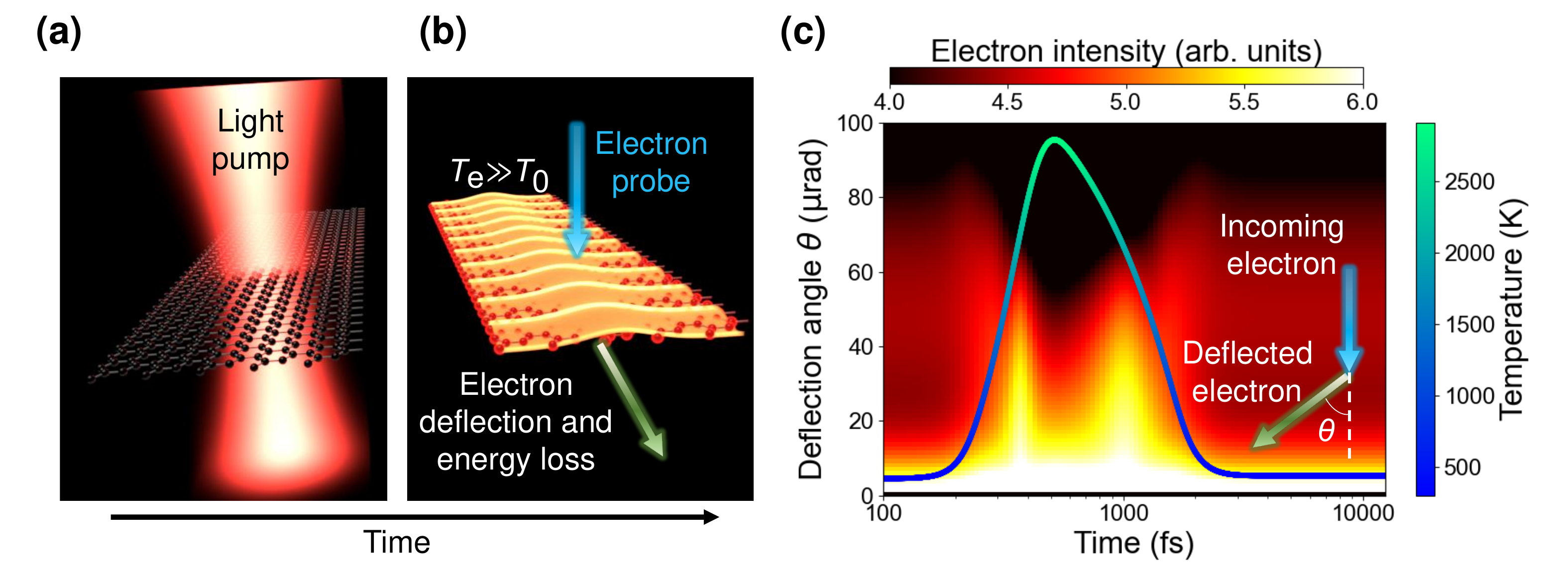}
\caption{Schematic illustration of ultrafast optical-pump/electron-probe (OPEP) sampling through energy-momentum transfer. (a) An ultrashort light pulse irradiates the sample at the initial time and elevates its electronic temperature. (b) A delayed electron pulse probes the thermal dynamics of sample excitations, which are revealed by the energy loss and lateral deflection that they produce on the electron, depending on the spatial distribution of those excitations. (c) Following optical pumping, the electron temperature (curve and right scale) first raises rapidly and then decays with time (lower horizontal scale), giving rise to a temporal variation in the inelastic electron scattering probability (density plot) with deflection angle (vertical scale), here represented for a 100\,nm wide graphene ribbon (0.2\,eV Fermi energy, 4\,meV damping) sampled by 100\,keV electrons that lose 0.2\,eV.}
\label{Fig1}
\end{figure*}

\section{The ultrafast optical-pump/electron probe (OPEP) approach}

The electron signal carries spectral information on excitations in the sample, and in addition, the angular distribution of inelastically scattered electrons reveals the spatial characteristics of those excitations. The acquisition of energy-momentum-resolved maps of transmitted electrons can directly yield dispersion curves of the sample modes \cite{PSV1975}. OPEP further adds temporal resolution, as we illustrate in Figure\ \ref{Fig1}. The sample (a graphene ribbon in this example) is optically pumped with an ultrafast laser (Figure\ \ref{Fig1}a), which creates an elevated electronic temperature in the material that is probed at a later time by a delayed electron pulse (Figure\ \ref{Fig1}a). Incidentally, the temperature rise occurs rather early because of the $\sim\Te^3$ scaling of the electronic heat with temperature (see Appendix\ \ref{appendixtwoTs}). The dynamics of rapid femtosecond heating followed by the subsequent picosecond cooling of graphene electrons is traced through the delay-dependent variations observed in the distribution of scattered electrons, which is represented in Figure\ \ref{Fig1}c for a fixed lost energy using the methods and analysis explained below. Additional plots analogous to Figure\ \ref{Fig1}c are presented in supplementary Figure\ \ref{fig:S1} for different values of the energy loss and for graphite ribbons.

\begin{figure*}
\centering
\includegraphics[width=0.7\textwidth]{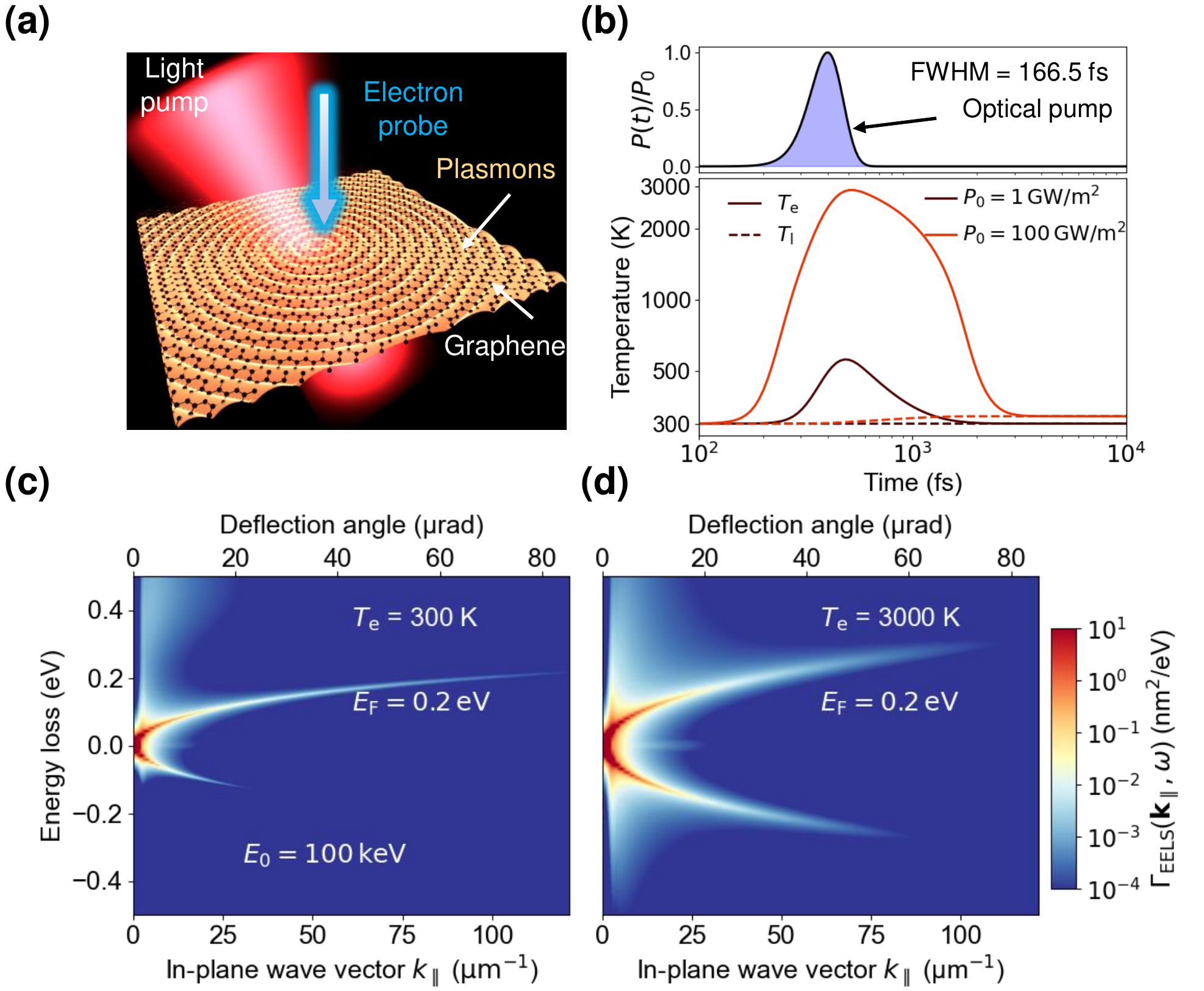}
\caption{OPEP characterization of plasmons in extended graphene. (a) Schematic representation of a graphene layer and OPEP configuration, with electrons impinging normal to the plane of the sample. (b) Temporal profiles of the optical-pump absorption (top panel, Gaussian 166.5\,fs FWHM, peaked at a time of 400\,fs) and sample temperatures (bottom panel). We plot the electron ($\Te$, solid curves) and lattice ($\Tl$, dashed curves) temperatures in the two-temperature model for two different normal-incidence peak absorption powers ($P_0=10^9$ and $10^{11}$\,W/m$^2$, pulse fluences of $(0.177/\pi\alpha)$ and $(17.7/\pi\alpha)$\,mJ/m$^2$, respectively, where $\pi\alpha\approx0.023$ is the absorbance of free-standing graphene in the visible range\cite{NBG08}). (c, d) Momentum- and energy-resolved loss probability $\Gamma_{\rm EELS}(\kparb,\omega)$ for 100\,keV transmitted electrons, revealing features associated with plasmon excitation in the sample for (c) low (300\,K) and (d) high (3000\,K) electronic temperature regimes. The graphene Fermi energy is $\EF=0.2\,$eV and the intrinsic damping is $\hbar\tau^{-1}=4\,$meV (lifetime $\tau=164.5\,$fs).}
\label{Fig2}
\end{figure*}


The power of momentum- and energy-resolved OPEP is illustrated in Figure\ \ref{Fig2} for a self-standing highly-doped extended graphene sample. Figure\ \ref{Fig2}a shows a scheme of the pump-probe configuration, with electrons incident normal to the graphene plane. When excited by an ultrashort optical pulse, high-energy electronic bands of graphene are populated, creating a nonequilibrium distribution of hot electrons, which quickly thermalizes to a high-temperature quasistationary state due to carrier-carrier scattering \cite{GSD08,GPM13}. During a subpicosecond timescale, the electronic temperature decreases as a result of a cascade of inelastic scattering processes, in particular by emitting and absorbing phonons \cite{HPB16}. Figure\ \ref{Fig2}b shows the temporal evolution of the temperature as modelled through the two-temperature model (see details in Appendix\ \ref{appendixtwoTs}) for two different optical pump fluences, reaching transient electronic temperatures as high as $\Te\sim3000\,$K. When probed with a delayed quasi-monochromatic electron pulse, the graphene plasmon dispersion can be mapped out from the energy- and angle-resolved inelastically scattered electron distribution. At room temperature (Figure\ \ref{Fig2}c), the dispersion relation is dominated by a plasmon band with a characteristic $\omega\sim\sqrt{\kpar}$ wave vector-frequency dispersion that is well documented for doped graphene \cite{paper235} (we use a Fermi energy $\EF=0.2\,$eV throughout this paper, see Appendix\ \ref{appendixEELS1} for details of the calculations). Interestingly, negative energy losses (i.e., energy gains) are observed from electrons that absorb thermally populated plasmons (Figure\ \ref{Fig2}c). Energy gains associated with optical phonons were equally observed in a pioneering experiment for electrons traversing thin LiF films \cite{BGS1966}, and more recently, this approach has been used to determine the phononic temperature in nanostructures \cite{LTH17,LB18}. In the present study, the gain dispersion band is resolved in momentum, showing mirror symmetry with respect to the horizontal axis, except for the difference in electron scattering probability, as losses are proportional to $n_\Te(\omega)+1$ and gains to $n_\Te(\omega)$, where $n_\Te(\omega)$ is the Bose-Einstein distribution at the electron temperature $\Te$ (see Appendix\ \ref{appendixEELS1}). At higher temperature (Figure\ \ref{Fig2}d) $n_\Te$ increases, thus reducing the relative difference between gain and loss probabilities. Additionally, the plasmon energy undergoes a clearly discernible blue shift because $\kB\Te$ ($\sim0.26\,$eV at $\Te=3000\,$K) exceeds $\EF=0.2\,$eV \cite{paper286}. We also observe an elevation in plasmon broadening beyond the intrinsic damping ($\hbar\tau^{-1}=4\,$meV) due to the availability of extra electron-hole-pair transitions that become accessible as $\Te$ increases \cite{paper235}. These conclusions are maintained when examining results for different values of $\EF$ (supplementary Figure\ \ref{fig:S2}) and multilayer graphene films (supplementary Figure\ \ref{fig:S3}). Incidentally, the fraction of inelastically scattering electrons is rather high at the relatively low plasmon energies under consideration \cite{paper228}, giving rise to plasmon replicas associated with multiple losses (supplementary Figure\ \ref{fig:S4}).

\begin{figure*}
\centering
\includegraphics[width=\textwidth]{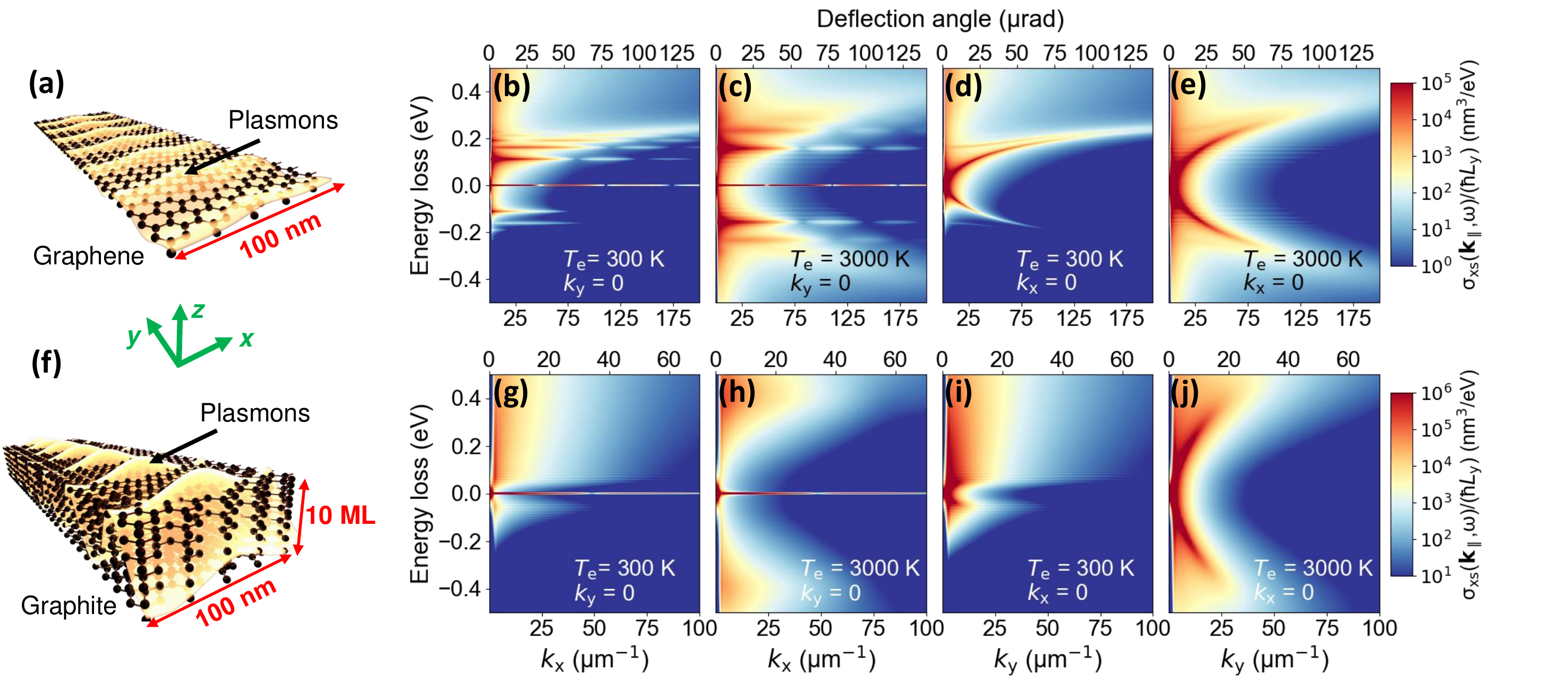}
\caption{Quantization of the plasmon lateral momentum in graphene and graphite ribbons characterized by OPEP. (a-e) We plot the energy- and momentum-resolved differential cross section of 100\,nm wide graphene ribbons ($0.2\,$eV Fermi energy, 4\,meV damping, see sketch in (a)) toward 100\,keV electron plane waves as a function of lost energy (vertical scales) and lateral momentum transfer (horizontal scales); we present cuts along both $k_x$ with $k_y=0$ (b,c) and $k_y$ with $k_x=0$ (d,e) for two different electronic temperatures $\Te$ (see labels). (f-j) Same as (a-e), but for 100\,nm wide, 3.3\,nm thick graphite ribbons (equivalent to 10 undoped graphene monolayers (MLs)) with the same intrinsic damping. We represent the momentum- and energy-resolved inelastic electron scattering cross section $\sigma_{\rm xc}(\kparb,\omega)$ normalized to the ribbon length $L_y$ (see Appendix\ \ref{appendixEELS2}).}
\label{Fig3}
\end{figure*}

\subsection{Revealing lateral plasmon confinement}

Ribbons break translational invariance and produce lateral plasmon confinement. We illustrate the resulting discretization in electron deflection in Figure\ \ref{Fig3}, where the momentum- and energy-resolved inelastic scattering cross section $\sigma_{\rm xc}(\kparb,\omega)$ is represented for an extended electron beam interacting with cool and heated graphene and graphite ribbons. This quantity is proportional to the loss probability, as explained in the Appendix\ \ref{appendixEELS2}. In Figure\ \ref{Fig3}a-e we show calculations for a 100\,nm wide graphene ribbon doped to $\EF=0.2\,$eV Fermi energy, whereas in Figure\ \ref{Fig3}f-j we consider a graphite ribbon of the same width and consisting of 10 monolayers (equivalent to $\approx3.3\,$nm thickness) of undoped graphene (i.e., we disregard any residual doping, which should be diluted in a larger number of layers). For graphene, the results for electron deflection in the plane containing the direction of the ribbon translational symmetry (Figure\ \ref{Fig3}d,e) are similar to those for planar graphene (Figure\ \ref{Fig2}), as expected from the similarity between the dispersion relations of the lowest-order monopolar plasmon waveguide in ribbons (see supplementary Figure\ \ref{fig:S5}) and plasmons in extended samples \cite{paper181}. The dipolar waveguide mode, which crosses $\kparb=0$ at finite energy $\sim0.2\,$eV, is also discernible, particularly at low temperature (Figure\ \ref{Fig3}d), while higher-order modes are not efficiently excited. In contrast, electron deflection along the transverse direction (Figure\ \ref{Fig3}b,c) exhibits sharp spectral features that reveal lateral confinement, accompanied by a milder momentum discretization resulting from the finite cosine-like charge distribution of plasmons across the ribbon. Like in extended graphene, an elevation in temperature produces plasmon blue shifts, an increase in spectral broadening, and a more symmetric gain-loss distribution. For graphite, the situation is different because the sample is undoped, so no plasmons are observed at low temperature (Figure\ \ref{Fig3}g,i), while a broad plasmon feature emerges at 3000\,K for electron deflection along the ribbon (Figure\ \ref{Fig3}j), which is quantized in energy for deflection across the ribbon (Figure\ \ref{Fig3}h), again due to lateral confinement. Additional plots offered in supplementary Figures\ \ref{fig:S6}, \ref{fig:S7}, and \ref{fig:S8} show the variation of the results in Figure\ \ref{Fig3} with graphene doping, graphene ribbon width, and graphite thickness, respectively.

\begin{figure*}
\centering
\includegraphics[width=0.9\textwidth]{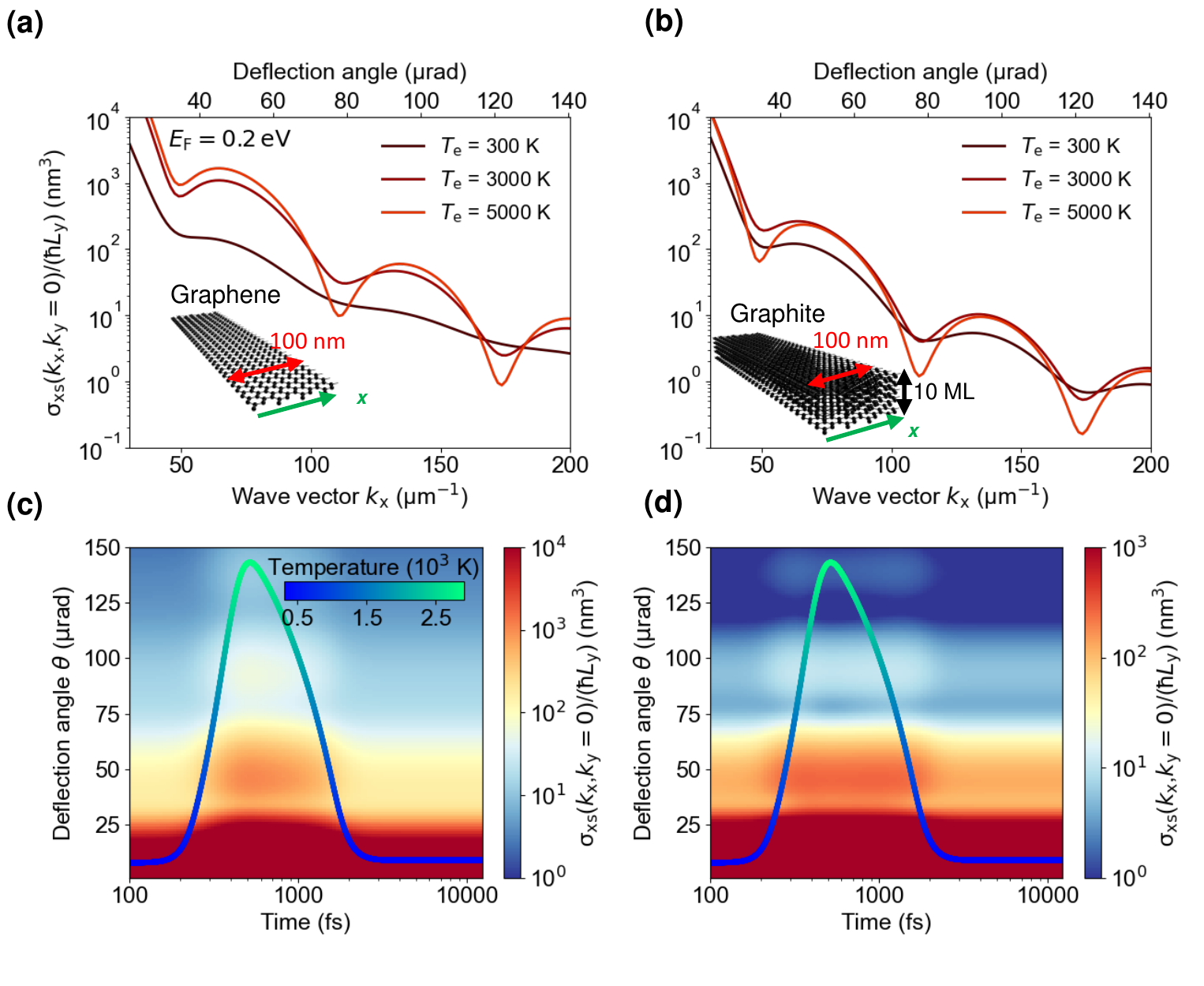}
\caption{Demonstration of spectrometer-free momentum-resolved OPEP. We plot the dependence on transverse momentum transfer (lower horizontal axes), or equivalently, deflection angle (upper axes), predicted for the energy-integrated, momentum-resolved inelastic electron cross section $\sigma_{\rm xc}(\kparb)=\int d\omega\;\sigma_{\rm xc}(\kparb,\omega)$ for (a,c) doped graphene and (b,d) undoped graphite samples with the same electron temperatures and additional parameters as in Figure\ \ref{Fig3}. The energy integral extends from -0.5\,eV to 0.5\,eV. We focus on $k_y=0$ (i.e., electrons scattered within a plane perpendicular to the ribbon axis) and show results (a,b) for selected temperatures and (c,d) for the temporal evolution under optical pumping, with the probability (density plots) and temperature (curves) shown in the same way as in Figure\ \ref{Fig1}b.}
\label{Fig4}
\end{figure*}

\subsection{Spectrometer-free momentum-resolved OPEP}

Quantization of momentum transfer due to lateral plasmon confinement suggests the possibility that these excitations and their temporal dynamics can be revealed by integrating the inelastic electron signal over a broad energy range, thus avoiding the need for highly-monochromatic electron beams and precise spectrometers. We explore this idea in Figure\ \ref{Fig4}a,b, where we present the energy-integrated (within the -0.5\,eV to 0.5\,eV range) cross sections extracted from Figure\ \ref{Fig3} for  electron deflection across the ribbon (i.e., as a function of $k_x$ for $k_y=0$). The results show clear momentum quantization in the inelastic electron signal, which becomes clearer as the temperature increases, particularly for the graphene ribbon. These observations suggest that the dynamics of the system could also be followed by measuring the energy-integrated electron angular distribution in the Fourier plane of the electron microscope. The delay-time dependence of the electron signal is shown in Figure\ \ref{Fig4}c,d (density plots), following the evolution of the electronic temperature (curves) in graphene and graphite ribbons upon optical pumping. Our calculations corroborate the increase in the visibility of the oscillations observed in the inelastic scattering probability as a function of deflection angle around the time of maximum heating.

\section{Conclusion}

Besides its fundamental interest, the study of ultrafast thermal dynamics of material excitations opens exciting opportunities for applications in optical switching and light modulation \cite{paper337}. In this work, we have demonstrated based on solid theoretical calculations that the optical-pump/electron-probe approach, which is becoming accessible within a growing number of ultrafast electron microscope setups, grants us access into nanoscale details of such dynamics combined with femtosecond temporal resolution. This method, which brings a radical enhancement in spatial resolution compared to alternative diffraction-limited optical probes, can rely on spectrometer-free momentum-resolved electron detection (i.e., in the microscope Fourier plane) when sampling nanostructures with a well-defined characteristic length, such as the width in ribbons, leading to momentum quantization due to lateral confinement of the supported excitations. In addition, the sampled mode energies can be arbitrarily low, provided their spatial extension is small enough to produce measurable transfers of lateral momentum to the electrons. We have illustrated the power of this concept by showing that energy-integrated, angle-resolved electron signals can reveal plasmons in structures of $\sim100\,$nm lateral size, which produce electron deflection angles that are sufficiently large to be resolved by a large fraction of existing transmission electron microscopes. The proposed approach should be generally applicable to study surface excitations in 2D materials, as well as local details of insulator-metal transition in vanadium and indium-titanium oxides, where the electron and lattice dynamics triggered by pumping with ultrafast laser pulses could provide information on collective electronic and vibronic excitations with combined nanometer and femtosecond spatiotemporal resolution.

\begin{widetext}

\section*{APPENDIX}
\appendix
\label{Methods}

\section{Electron energy-loss and -gain probabilities in extended planar films}
\label{appendixEELS1}

We follow a well-established formalism \cite{paper228} to calculate the loss probability of an electron that is normally impinging on a planar thin film. The distribution of transmitted electrons as a function of transferred transverse momentum $\hbar\kparb$ and energy $\hbar\omega$ is given by
\begin{align}
\Gamma_{\rm EELS}^0(\kparb, \omega)=\dfrac{2e^2}{\hbar \pi^2 v^2} \dfrac{k_{\parallel}}{(k_{\parallel}^2 + \omega^2/v^2)^2} {\rm Im}\{\rp(\kpar, \omega)\}
\nonumber
\end{align}
at zero electronic temperature, where we disregard small retardation effects for simplicity. Here, $\rp(\kpar,\omega)$ denotes the momentum- and frequency-dependent Fresnel reflection coefficient of the film for p polarization, which is in turn expressed in terms of the surface conductivity $\sigma(\kpar,\omega)$ (see Appendix\ \ref{appendixsigma}) as
\begin{align}
\rp(\kpar, \omega) = \dfrac{1}{1-\ii \omega/[2\pi\kpar\,\sigma(\kpar,\omega)]},
\nonumber
\end{align}
where we neglect retardation effects. We thus describe the film as a zero-thickness layer under the assumption that the involved surface modes have long wavelengths compared to the film thickness. For graphene, we use the random-phase approximation (RPA) conductivity (see Appendix\ \ref{appendixsigma}), while a thin graphite film consisting of $N$ graphene planes is represented by the graphene conductivity multiplied by $N$. For finite electronic temperature $\Te$, the electron energy-loss probability needs to be corrected due to the following two effects: (1) the reflection coefficient is modified by the thermal dependence of the film conductivity (see Appendix\ \ref{appendixsigma}); and (2) the thermal population of excited electronic states in the film produces an increase in energy losses, as well as a finite probability of energy gains, captured by the expression \cite{LS1972,LK1970}
\begin{align}
\Gamma_{\rm EELS}^\Te(\kparb,\omega) = \Gamma_{\rm EELS}^0(\kparb,|\omega|)\;[n_\Te(\omega)+1]\;\left[\Theta(\omega)-\Theta(-\omega)\right],
\nonumber
\end{align}
where 
\begin{align}
n_\Te(\omega)=\dfrac{1}{\ee^{\hbar\omega/\kB \Te} - 1}
\nonumber
\end{align}
is the frequency- and temperature-dependent Bose-Einstein distribution, the $\Theta(-\omega)$ term accounts for energy gain, and we have used the property $-n_\Te(-\omega)=n_\Te(\omega)+1$. In the present work, we apply this model to describe inelastic electron scattering in graphene and few-layer graphite extended films.

\section{Inelastic electron scattering cross section of planar nanostructures}
\label{appendixEELS2}

We consider a free electron moving along $z$ and initially prepared in a plane wave state $\psi_i(\rb)=\ee^{\ii p_{i,z}z}/\sqrt{AL}$ of energy $\hbar\varepsilon_i$ and momentum $\hbar p_{i,z}\zz$, where $A$ and $L$ are the transverse area and longitudinal length of the quantization box, respectively. The electron is taken to interact with a planar structure lying in the $z=0$ plane. We aim to calculate the transition probability to a final state $\psi_f(\rb)=\ee^{\ii p_{f,z}z}\ee^{\ii\kparb\cdot\Rb}/\sqrt{AL}$ of energy $\hbar\varepsilon_f$ and momentum $\hbar(\kparb+p_{f,z}\zz)$, where $\kparb$ is the transverse wave vector transfer and $\Rb=(x,y)$. Neglecting retardation, the energy-resolved inelastic transition rate is given by \cite{paper149}
\begin{align}\nonumber
\dfrac{d \Gamma(\omega)}{dt}=\dfrac{2e^2}{\hbar} \sum_f\int d^3\rb\int d^3\rb'\;\psi_f(\rb)\psi_i^{*}(\rb)\psi_f^*(\rb')\psi_i(\rb')\,{\rm Im}\{-W^{\rm ind}(\rb,\rb',\omega)\}\,
\delta(\varepsilon_f-\varepsilon_i+\omega), \nonumber
\end{align}
where $W^{\rm ind}(\rb,\rb',\omega)$ is the induced part of the screened Coulomb interaction. We now (1) make the substitution $\sum_f\rightarrow AL\,(2\pi)^{-3}\int d^2\kparb\int dp_{f,z}$, (2) adopt the nonrecoil approximation to express the transition frequency as $\varepsilon_i-\varepsilon_f\approx v(p_{i,z}-p_{f,z})$, where $v$ is the electron velocity, and (3) divide the rate $d\Gamma/dt$ by the incident electron current density $v/(AL)$ to obtain the spectrally-resolved inelastic scattering cross section $\sigma_{\rm xc}(\omega)=\int d^2\kparb\;\sigma_{\rm xc}(\kparb,\omega)$, where
\begin{align}
\sigma_{\rm xc}(\kparb,\omega)=\frac{e^2}{4\pi^3\hbar v^2}\int d^3\rb\int d^3\rb'
\;\ee^{\ii\kparb\cdot(\Rb-\Rb')}\ee^{\ii\omega(z'-z)/v}
\;{\rm Im}\{-W^{\rm ind}(\rb,\rb',\omega)\} \label{sigmakparw}
\end{align}
is the cross section resolved in momentum and energy transfers $\hbar\kparb$ and $\hbar\omega$.

We describe the planar structure through a local surface conductivity $\sigma(\omega)\equiv\sigma(\kpar=0,\omega)$. Using a quasistatic eigenmode expansion detailed elsewhere \cite{paper228}, this allows us to express the screened interaction as
\begin{align}
W^{\rm ind}(\rb,\rb',\omega)=D\sum_j\dfrac{1}{1/\eta_j-1/\eta(\omega)}\,\phi_j(\rb)\phi_j(\rb')
\label{Wind}
\end{align}
in terms of size-independent real-valued charge distributions $\rho_j(\thetav)$ and eigenvalues $\eta_j$, where $\eta(\omega)=\ii\sigma(\omega)/\omega D$,
\begin{align}
\phi_j(\rb)=\int d^2\thetav\;\dfrac{\rho_j(\thetav)}{|\rb-D\thetav|}
\label{phij}
\end{align}
are normalized scalar mode potentials, $\thetav=\Rb/D$ is the in-plane position vector $\Rb$ normalized to the ribbon width $D$, and $j$ labels different modes (see below). Inserting Eq.\ (\ref{Wind}) into Eq.\ (\ref{sigmakparw}), we find
\begin{align}
\sigma_{\rm xc}(\kparb,\omega)=\frac{e^2D}{4\pi^3\hbar v^2}\sum_j {\rm Im}\left\{\dfrac{1}{1/\eta(\omega)-1/\eta_j}\right\}\left|\tilde{\phi}_j(\kb)\right|^2
\label{sigmabis}
\end{align}
where $\tilde{\phi}_j(\kb)=\int d^3\rb\,\ee^{\ii\kb\cdot\rb}\phi_j(\rb)$ and $\kb=\kparb+(\omega/v)\zz$. Then, using the Fourier transform $4\pi/k^2$ of the Coulomb potential $1/r$ in Eq.\ (\ref{phij}), we find $\tilde{\phi}_j(\kb)=(4\pi/k^2)\tilde{\rho}_j(\kparb)$, where
\begin{align}
\tilde{\rho}_j(\kparb)=\int d^2\thetav\;\rho_j(\thetav)\,\ee^{-\ii\kparb\cdot\thetav D}.
\label{tilderho}
\end{align}
This allows us to recast Eq.\ (\ref{sigmabis}) as
\begin{align}
\sigma_{\rm xc}(\kparb,\omega)=\frac{4e^2D}{\pi\hbar v^2}\,\frac{1}{\big(\kpar^2+\omega^2/v^2\big)^2}\sum_j {\rm Im}\left\{\dfrac{1}{1/\eta(\omega)-1/\eta_j}\right\}\left|\tilde{\rho}_j(\kparb)\right|^2,
\label{sigmafinal}
\end{align}
which has units of time$\times$(length)$^4$.

Here, we apply this formalism to ribbons of width $D$ that possess translational invariance along $y$ (Figure\ \ref{Fig3}a,b), so it is convenient to multiplex the mode index as $j\rightarrow\{j,q\}$ into a transverse index (we retain $j$ for this purpose) and a wave vector $q/D$ along $y$. This needs to be accompanied by the substitutions $\sum_j\rightarrow\sum_{jq}$ and $\rho_j(\thetav)\rightarrow\rho_{jq}(\theta_x)\ee^{\ii q\theta_y}\sqrt{D/L_y}$, where $L_y\gg D$ is the ribbon length and we incorporate the wave-plane dependence on $\theta_y$ in the charge distribution. We now rewrite Eq.\ (\ref{tilderho}) as $\tilde{\rho}_j(\kparb)\rightarrow\tilde{\tilde{\rho}}_j(\kparb)\,\delta_{q,k_yD}\,\sqrt{L_y/D}$ by making the reassignment
\begin{align}
\tilde{\tilde{\rho}}_j(\kparb)=\int_{-1/2}^{1/2} d\theta_x\;\rho_{j,k_yD}(\theta_x)\,\ee^{-\ii\theta_yk_yD}.
\label{tilderhoribbon}
\end{align}
Finally, the counterpart of Eq.\ (\ref{sigmafinal}) for translationally invariant ribbons reduces to
\begin{align}
\dfrac{\sigma_{\rm xc}(\kparb,\omega)}{L_y}&=\frac{4e^2}{\pi\hbar v^2}\,\frac{1}{\big(\kpar^2+\omega^2/v^2\big)^2}\sum_j {\rm Im}\left\{\dfrac{1}{1/\eta(\omega)-1/\eta_{j,k_xD}}\right\}\left|\tilde{\tilde{\rho}}_j(\kparb)\right|^2,
\label{sigmafinalribbon}
\end{align}
which is normalized to the ribbon length $L_y$ and has units of time$\times$(length)$^3$. We use Eqs.\ (\ref{tilderhoribbon}) and (\ref{sigmafinalribbon}) to calculate the results presented in Figures\ \ref{Fig3} and \ref{Fig4}.

\section{Graphene conductivity at finite temperature}
\label{appendixsigma}

The temperature-dependent nonlocal RPA surface conductivity of graphene is given by \cite{WSS06,HD07}
\begin{align}
\sigma^0(\kpar,\w) = \dfrac{\ii e^2\omega}{2\pi^2\hbar k_{\parallel}^2}\int d^2 \qv\sum_{s,s'=\pm1}\left[1 + ss'\dfrac{\qv\cdot(\kparb+\qv)}{q\abs{\kparb+\qv}}\right]\dfrac{\nF(s'\hbar\vF\abs{\kparb + \qv})-\nF(s\hbar\vF q)}{\w - \vF(sq-s'\abs{\kparb+\qv})+\ii 0^{+}},
\label{NonzeroT_Chi}
\end{align}
where the superscript 0 indicates that inelastic relaxation occurs at an infinitesimal rate, $\vF\approx10^6\,$m/s is the Fermi velocity, $\Te$ is the electronic temperature, $\nF(E) =\left[\ee^{(E-\mu)/\kB\Te}+1\right]^{-1}$ is the Fermi-Dirac distribution function, and $\mu$ is the temperature-dependent chemical potential. The latter can be approximated as \cite{paper286} $\mu\approx\EF\left[\left(1+\xi^2\right)^{1/2}-\xi\right]^{1/2}$, where $\EF=\hbar\vF\sqrt{\pi n}$ is the zero-temperature Fermi energy for a doping carrier density $n$, and $\xi=(2\log^24)(\kB\Te/\EF)^2$. At $\Te=0$, we have $\nF(E)=\Theta(\EF-E)$ and the conductivity admits the analytical expression \cite{WSS06,HD07}
\begin{align}
\sigma^0_{\Te=0}(k_{\parallel},\w,\EF) = \dfrac{-2\ii e^2\omega}{\pi\hbar\vF}\left[\dfrac{\kF }{\kpar^2} + \dfrac{[G(\Delta_{-})-G(-\Delta_{-})+\ii\pi]\Theta({\rm Re}\{\Delta_{-}\}+1)+G(-\Delta_{-})-G(\Delta_{+})}{8\sqrt{\w^2/\vF^2-k_{\parallel}^2}}\right],
\nonumber
\end{align}
with $G(z) = z \sqrt{z^2-1} - \log(z + \sqrt{z^2 - 1})$ and $\Delta_{\pm} = (\w/\vF\pm 2\kF)/\kpar$, where the square root is taken to yield positive real parts and the imaginary part of the log function is taken in the $[-\pi,\pi)$ sheet. As a more efficient alternative to evaluating the integral in Eq.\ (\ref{NonzeroT_Chi}), we calculate the $\Te$-dependent conductivity from the $\Te=0$ expression using the identity \cite{M1978}
$\nF(E)=(4\kB \Te)^{-1}\int_{-\infty}^{\infty}dx\,\Theta(x - E)\big{/}{\rm cosh}^2\left[(\mu-x)/(2\kB\Te)\right]$, which allows us to write
\begin{align}
\sigma^0(\kparb,\w)=\dfrac{1}{4\kB\Te}\int_{-\infty}^{\infty}dx\;\sigma^0_{\Te=0}(\kpar,\w,x)\;{\rm cosh}^{-2}\!\left(\dfrac{\mu-x}{2\kB\Te}\right).
\nonumber
\end{align}
Finally, we introduce a phenomenological inelastic lifetime $\tau$ using the Mermin prescription \cite{M1970}
\begin{align}
\sigma(\kparb,\w)=\dfrac{(1+\ii/\w\tau)\sigma^0(\kparb,\w+\ii/\tau)}{1+(\ii/\w\tau)\sigma^0(\kparb,\w+\ii/\tau)/\sigma^0(\kparb,0)},
\nonumber
\end{align}
which is designed to preserve the local electron density. Although $\tau$ exhibits a complex dependence on temperature (see supplementary Figure\ \ref{fig:S9}a,b), we adopt for simplicity a constant value given by $\hbar\tau^{-1}=4\,$meV throughout this work. Finally, we note that we describe graphite films consisting of $N$ undoped graphene layers by means of a temperature-dependent surface-conductivity $N\sigma(\kparb,\w)$ evaluated at $\EF=0$.

\section{Two-temperature model}
\label{appendixtwoTs}

We ignore thermal diffusion under the assumption that the structures are pumped with spatially homogeneous illumination and further neglect radiative emission in our self-standing samples. Then, for simplicity, we find the time- and position-dependent electronic and lattice temperatures, $\Te(\Rb,t)$ and $\Tl(\Rb,t)$, by solving a stripped version of the two-temperature model equations \cite{AKP1974}
\begin{align}
 &\ce \frac{d\Te}{dt}=p^{\rm abs}+A(\Te^3-\Tl^3), \nonumber\\
 &\cl \frac{d\Tl}{dt}=-A(\Te^3-\Tl^3),
\nonumber
\end{align}
where $\ce$ and $\cl$ are the graphene electronic and phononic heat capacities per unit area, $p^{\rm abs}(\Rb,t)$ is the absorption power density due to optical pumping, and the rightmost terms account for electron-phonon coupling. We assume that the latter is dominated by disorder, which leads to the $\sim T^3$ scaling \cite{SRL12,paper313} with $A=(1.2\mathcal{D}^2|\mu|\kB^3)/(\pi^2 \rho \hbar^4 \vF^3 v_s^2 L)$, where $\rho=7.6\times 10^{-8}~\mathrm{g/cm^2}$ and $\mathcal{D}\approx 40~\mathrm{eV}$ are the graphene mass density and deformation potential, respectively, $v_s\approx0.02\,\vF$ is the graphene sound velocity, and $L\approx10\,$nm. The electronic heat capacity $\ce=\partial Q_{\rm e}/\partial \Te$ is obtained as the derivative of the surface heat density \cite{paper286} $Q_{\rm e}=\beta(\kB \Te)^3/(\hbar \vF)^2$, where $\beta=(2/\pi)\int_{0}^{\infty} x^2 dx \left[(\ee^{x+\mu/\kB \Te}+1)^{-1} + (\ee^{x-\mu/\kB \Te}+1)^{-1} \right] - (2/3\pi)(\EF/\kB\Te)^3=-(4/\pi)[{\rm Li}_3(-\ee^{-\mu/\kB\Te})+{\rm Li}_3(-\ee^{\mu/\kB\Te})]-(2/3\pi)(\EF/\kB\Te)^3$, where ${\rm Li}_n(x)=\sum_{k=1}^{\infty} x^k/k^n$ is the polylogarithm function of order $n$. The phonon heat capacity is calculated as \cite{BLC96} $\cl=9k_{\rm B}^3\Tl^2\zeta(3)/(\pi\hbar^2v_{\rm ph}^2)$, where $v_{\rm ph}\approx 10^4\,$m/s is the phonon velocity and $\zeta(3)\approx1.202$ is the Riemann Zeta function; this expression is valid for small $\Tl$ compared with the Debye temperature $\sim1000\,$K in graphene (see Figure\ \ref{Fig1}b). Actually, $\cl$ is several orders of magnitude higher than $\ce$ (see supplementary Figure\ \ref{fig:S9}c), which implies that the former plays a minor role and $\Tl$ does not increase substantially compared with $\Te$ (see Figure\ \ref{Fig2}b).

\section{Summary of quasistatic eigenmodes for ribbons}
\label{summaryche}

We consider a ribbon of width $D$ having translational invariance along $y$, for which we intend to find mode eigenvalue and eigenfunctions $\eta_s$ and $\tilde{\rho}_j(\kparb)$, where $\kparb=(k_x,k_y)$ and we use the combined mode index $s=\{j,q=k_yD\}$. This problem has been addressed using different methods, including electromagnetic simulations \cite{paper176}, direct solution of the associated self-consistent quasistatic integral equation \cite{SSN11,TFB89}, and inversion of the corresponding integral eigenvalue problem in real-space and special-function representations of the mode fields \cite{paper228,KR14,paper303,GSC20}. Here, we find fast convergence in the solution of the eigenvalue problem by using a Chebyshev expansion of the  electric field, as shown in detail in Appendix\ \ref{sec:Chebyshev}, which yields the following result for the Fourier transform of the mode charge density (see Eq.\ (\ref{tilderhoribbon})):
\begin{align}
\tilde{\tilde{\rho}}_j(\kparb)=\dfrac{\pi^2}{2}\sum_{n=0}^{\infty}(-\ii)^{n-1}\left\{4(n+1)u_{s,n}J_{n+1}\left(\dfrac{k_x D}{2}\right)-\ii k_y D v_{s,n}\left[J_{n}\left(\dfrac{k_x D}{2}\right) +J_{n+2}\left(\dfrac{k_x D}{2}\right)\right]\right\},
\nonumber
\end{align}
where $J_n$ is a Bessel function of order $n$, and $u_{s,n}$, $v_{s,n}$, and $\eta_s$ (see tabulated values in supplementary Figure\ \ref{fig:S5}a and Tables\ \ref{tab:eta} and \ref{tab:uvj:q0}) are determined from the eigensolutions of the $2N\times2N$ matrix equation
\begin{align}
 \dfrac{1}{\eta_s} 
 \begin{bmatrix}
\mathcal{A} & 0 \\
0 & \mathcal{A}
\end{bmatrix} \cdot
\begin{bmatrix}
    \mathbf{u}_s \\ \mathbf{v}_s
\end{bmatrix}
= 
\begin{bmatrix}
 \mathcal{M}^{11} & \mathcal{M}^{12}\\
 \mathcal{M}^{21} & \mathcal{M}^{22}
 \end{bmatrix}
 \cdot
\begin{bmatrix}
    \mathbf{u}_s \\ \mathbf{v}_s
\end{bmatrix},
\label{eq:matrixeq}
\end{align}
with $\mathbf{u}_s=\big[u_{s,0} \cdots u_{s,N-1} \big]^{\rm T}$ and $\mathbf{v}_s=\big[v_{s,0}\cdots v_{s,N-1}\big]^{\rm T}$ (T stands for transpose). Matrices in Eq.\ (\ref{eq:matrixeq}) are defined in terms of $N\times N$ blocks with coefficients
\begin{align}
\mathcal{A}_{ij} &= U_{j}(t_{i})\sqrt{1 - t_{i}^2}, \nonumber\\
\mathcal{M}^{11}_{ij} &= 4\sum_{m=0}^{\infty}\left[ D_{m}^{(0)}(0,q,t_i)F_{mj}^{(1)}(t_i) - 2D_{m}^{(1)}(0,q, t_i)F_{mj}^{(2)}(t_i) - D_{m}^{(2)}(0,q, t_i)F_{mj}^{(3)}(t_i) + K_{m}^{(2)}(q,t_i)F_{mj}^{(4)}\right], \nonumber\\
\mathcal{M}^{12}_{ij} &= -\mathcal{M}^{21}_{ij}= 2q\sum_{m=0}^{\infty}\left[ -D_{m}^{(0)}(0,q,t_i)F_{mj}^{(2)}(t_i) - D_{m}^{(1)}(0,q,t_i)F_{mj}^{(3)}(t_i)+ K_{m}^{(1)}(q,t_i)F_{mj}^{(4)}\right], \nonumber\\
\mathcal{M}^{22}_{ij} &= -q^2\sum_{m=0}^{\infty}\left[-D_{m}^{(0)}(0,q,t_i)F_{mj}^{(3)}(t_i)+ K_{m}^{(0)}(q,t_i)F_{mj}^{(4)}\right],
\nonumber
\end{align}
where $t_i=\cos\left[\pi(i+1)/(N+1)\right]$, the indices $i$ and $j$ run from 0 to $N-1$,
\begin{align}
K_m^{(n)}(q,t) &= [\log(q/4)+\gamma]\,D^{(n)}_{m}(0,q,t)+2\sum_{k=1}^{\infty}\frac{1}{k}D^{(n)}_{m}(k,q,t), \quad\quad \text{with}\;n=0,1,2, \nonumber\\
D^{(0)}_{m}(l,q,t)&=\sum_{k = -\infty}^{\infty}(-1)^k C_{m}(k+2l,q) I_{k}(qt/2), \nonumber\\
D^{(1)}_{m}(l,q,t)&=-\dfrac{q}{4}\sum_{k = -\infty}^{\infty}(-1)^k \left[C_m(k+2l+1,q)+C_m(k+2l-1,q)\right] I_{k}(qt/2), \nonumber\\
D^{(2)}_{m}(l,q,t)&=\dfrac{q^2}{16}\sum_{k = -\infty}^{\infty}(-1)^k \left[C_m(k+2l+2,q)+2C_m(k+2l,q)+C_m(k+2l-2,q)\right] I_{k}(qt/2), \nonumber\\
C_n(\nu,q)&=\ii^{-\nu}(2 - \delta_{n0})I_{(n+\nu)/2}\left(q/4\right)I_{(n-\nu)/2}\left(q/4\right) \times\left\{\begin{matrix}
\cos{(\pi\nu/2)},\quad\quad\text{for even $n$}, \\
\ii\,\sin{(\pi\nu/2)},\quad\quad\text{for odd $n$},
\end{matrix}\right.
\nonumber
\end{align}
$I_{n}$ is a modified Bessel function of order $n$, 
\begin{align}
F^{(1)}_{mp}(t) =& -\dfrac{\pi}{2}\left[(p+m+1)U_{p+m}(t) + (p-m+1)U_{\abs{p-m+1}-1}(t)\right], \nonumber\\
F^{(2)}_{mp}(t) =& \dfrac{\pi}{2}\left[T_{p+m+1}(t) + (-1)^{p-m+1} T_{\abs{p-m+1}}(t)\right], \nonumber\\
F^{(3)}_{mp}(t) =&\dfrac{-\pi}{4}\bigg[L_{p+m}(t)+L_{\abs{m-p}}(t) - L_{m+p+2}(t) - L_{\abs{m-p-2}}(t)\bigg], \nonumber\\
F^{(4)}_{mp} =& \dfrac{\pi}{4}(1-\delta_{m0})\left(\delta_{mp} - \delta_{mp+2}\right), \nonumber
\end{align}
$T_n(t)$ and $U_n(t)$ are Chebyshev polynomials (defined by $T_n(\cos \theta)=\cos(n\theta)$ and $U_n(\cos \theta)=\sin[(n+1)\theta]/\sin\theta$), $L_0(t)=T_0(t)\log2$, and $L_{m>0}=T_m(t)/m$.

The eigenvectors $\mathbf{u}_s$ and $\mathbf{v}_s$ must be normalized in such a way that the mode electric fields satisfy $\int_{-1/2}^{1/2} d\theta_x \left[ \mathcal{E}_{sx}(\theta_x) \mathcal{E}_{s'x}^*(\theta_x) + \mathcal{E}_{sy}(\theta_x) \mathcal{E}_{s'y}^*(\theta_x)\right] = \delta_{ss'}$ (with $q=q'$), where $\theta_x=x/D$ and
\begin{align}
\mathcal{E}_{sx}(\theta_x)&= \sum_{n=0}^{\infty} u_{s,n} \sqrt{1-4\theta_x} U_n(2\theta_x), \nonumber\\
\mathcal{E}_{sy}(\theta_x)&= -\ii \sum_{n=0}^{\infty} v_{s,n} \sqrt{1-4\theta_x} U_n(2\theta_x) \nonumber
\end{align}
are the $x$ and $y$ components of the electric field of mode $s=\{j,q\}$.

\section{Detailed derivation of a solution of quasistatic ribbon plasmon eigenfunctions through the Chebyshev expansion method}\label{sec:Chebyshev}

We use a Chebyshev polynomial expansion of the optical electric field to calculate semi-analytically the plasmonic eigenmodes of a graphene ribbon of width $D$ lying on the $z=0$ plane and having translational invariance along $y$. The ribbon is taken to occupy the $-D/2<x<D/2$ region. We describe graphene by means of a local, frequency-dependent surface conductivity $\sigma(\w)$ and incorporate the dependence on surface position $\Rv=(x,y)$ by writing $\sigma(\Rv,\w) = \sigma(\w)f(\Rv)$, where $f(\Rv)=1$ if $|x|\le D/2$ and 0 otherwise. The monochromatic optical electric field $\Eb(\Rv,\w)$ in the graphene plane then satisfies the integral equation \cite{paper228}
\begin{equation}\label{Integral_equation_inRspace}
\Ev(\Rv,\w)= \Ev^{\rm ext}(\Rv,\w) +
\dfrac{\ii\sigma(\w)}{\bar{\eps}\,\w}\nabla_{\Rv}\int \dfrac{d^2\Rv'}{\abs{\Rv-\Rv'}}\nabla_{\Rv'}f(\Rv')\cdot\Ev(\Rv', \w),
\end{equation}
where $\bar{\epsilon}$ is the average permittivity of the embedding medium (see below). Following Ref.\ \citenum{paper228}, we define dimensionless coordinates $\thetav = \Rv/D$ and the normalized electric field $\Eepsv(\thetav) = D\sqrt{f(\thetav)}\Ev$ to rewrite Eq.~\eqref{Integral_equation_inRspace} as
\begin{align}
\Eepsv(\thetav,\w)&= \Eepsv^{\rm ext}(\thetav,\w) +
\eta(\omega)\int d^2\thetav'\;\mathbf{M}(\thetav, \thetav')\cdot\Eepsv(\thetav', \w),
\nonumber
\end{align}
where
\begin{align}
\mathbf{M}(\thetav, \thetav')&= \sqrt{f(\thetav)f(\thetav')}\;\nabla_{\thetav}\otimes\nabla_{\thetav}\;\dfrac{1}{\abs{\thetav-\thetav'}}
\nonumber
\end{align}
and \[\eta(\w) = \frac{\ii\sigma(\omega)}{D\omega\eps}.\] In the absence of an external field, the above equation reduces to an eigenvalue problem:
\begin{equation}
\Eepsv_{j}(\thetav,\w)= \eta_{j}\int d^2\thetav'\;\mathbf{M}(\thetav, \thetav')\cdot\Eepsv_{j}(\thetav',\w).
\label{eq:genexp}
\end{equation}
Since the kernel $\mathbf{M}$ is real and symmetric, we can find a complete set of orthonormal solutions $\Eepsv_{j}$ that satisfy
\begin{align}
\int d^2\thetav\; \Eepsv_{j}(\thetav)\cdot \Eepsv^*_{j'}(\thetav) &= \delta_{jj'},
\nonumber\\
\sum_{j} \Eepsv_{j}(\thetav)\otimes\Eepsv_{j}(\thetav') &= \delta(\thetav-\thetav')\II_{2 \times 2}, \nonumber
\end{align}
where $\II_{2 \times 2}$ is the $2\times2$ unit matrix.

We now focus on the specific geometry of a graphene ribbon of width $D$. Considering its translational invariance along $y$, we can multiplex the mode index $j$ into a normalized wave vector $q=k_yD$ and the mode order for each fixed value of $q$ (we also use $j$ for the mode order). The spatial dependence of mode $s\equiv\{j,q\}$ is thus given by 
\begin{equation}
\Eepsv_{jq}(\theta_x)\,\ee^{\ii q \theta_y}.
\nonumber
\end{equation}
Using this, Eq.\ \eqref{eq:genexp} can be recast as
\begin{equation}
\Eepsv_{s}(\theta_x) = 2\eta_{s}\int_{-1/2}^{1/2} d\theta^{\prime}_x\,(\nabla_{\thetav}\otimes\nabla_{\thetav})K_{0}\big(q\abs{\theta_x-\theta^{\prime}_x}\big)\,\cdot\Eepsv_{s}(\theta^{\prime}_x),
\label{ieK0}
\end{equation}
where $\nabla_{\thetav}=\partial_{\theta_{x}}\xx+\ii q\yy$ and we have made use of the identity
\begin{equation}
\int d\theta^{\prime}_y\,\ee^{\ii q\theta_y}/|\thetav-\thetav'|= 2K_{0}\big(q|\theta_x-\theta^{\prime}_x|\big).
\nonumber
\end{equation}
Now, the integral equation (\ref{ieK0}) can be written in the form 
\begin{equation}
\dfrac{1}{\eta_{s}}\begin{bmatrix}
\Eeps_{sx}(\theta_x)\\
\ii{\Eeps}_{sy}(\theta_x)
\end{bmatrix} 
 = 2\int_{-1/2}^{1/2} d\theta_{x}^{\prime}\begin{bmatrix}
 \partial_{\theta_{x}}^2 & -q\partial_{\theta_{x}}\\
 q\partial_{\theta_{x}} & -q^2
 \end{bmatrix}
 K_{0}(q\abs{\theta_{x}-\theta_{x}^{\prime}})\begin{bmatrix}
\Eeps_{sx}(\theta_{x}^{\prime})\\
\ii{\Eeps}_{sy}(\theta_{x}^{\prime})
\end{bmatrix}.
\nonumber
\end{equation}
To apply the Chebyshev expansion method, it is convenient to map the integration domain onto the $[-1, 1]$ interval by introducing the variable changes $2\theta_{x} = t$ and $2\theta_{x}^{\prime} = t'$:
\begin{equation}
\label{eq:mateq1}
\dfrac{1}{\eta_{s}}\begin{bmatrix}
\Eeps_{sx}(t)\\
\ii{\Eeps}_{sy}(t)
\end{bmatrix}
 =4\int_{-1}^{1} dt'\begin{bmatrix}
 \partial_{t}^2 & (q/2)\partial_{t}\\
 -(q/2)\partial_{t} & -(q/2)^2
 \end{bmatrix}
K_{0}\left(q\abs{t-t'}/2\right)\begin{bmatrix}
\Eeps_{sx}(t')\\
\ii{\Eeps}_{sy}(t')
\end{bmatrix}.
\end{equation}
The essence of the Chebyshev method lies on the expansion of the kernel function $K_{0}\left(q\abs{t-t'}/2\right)$ in terms of the Chebyshev polynomials $T_n(t)$ and $U_{n}(t)$, defined such that $T_n(\cos \theta)=\cos(n\theta)$ and $U_n(\cos \theta)=\sin[(n+1)\theta]/\sin\theta$  \cite{B01_2}. In order to do so, we recall that the modified Bessel function $K_0$ can be expanded into the Neumann series \cite{OLB10}
\begin{equation}
K_{0}(z) = -\left[\ln{(z/2)} + \gamma\right]I_{0}(z) +\sum_{k=1}^{\infty}\dfrac{2}{k}I_{2k}(z),
\nonumber
\end{equation}
where $\gamma\approx0.57721$ is the Euler constant and $I_{n}$ denotes the modified Bessel function of order $n$. In addition, one can use the Neumann addition formula for even-order $I_{2l}$
\begin{align}
I_{2l}(q\abs{t-t'}/2) = \sum_{k = -\infty}^{\infty} (-1)^{k} I_{2l+k}(q t'/2)I_{k}(q t/2)
\nonumber
\end{align}
to represent the kernel function as a separable product of functions with arguments $t'$ and $t$. The other kernel functions in Eq.\ \eqref{eq:mateq1} can be obtained by taking the first- and second-order derivatives of the above identity with respect to $t$:
\begin{align}
    \dfrac{\partial}{\partial t} I_{2l}(q\abs{t-t'}/2) 
    &= -\dfrac{q}{4}\sum_{k = -\infty}^{\infty} (-1)^{k+1}\left[I_{k+2l -1}(q t'/2) + I_{k + 2l-1}(q t'/2)\right]I_{k}(q t/2),
    \nonumber\\
    \dfrac{\partial^2}{\partial t^2} I_{2l}(q\abs{t-t'}/2) &= \dfrac{q^2}{16}\sum_{k = -\infty}^{\infty} (-1)^{k}\left[I_{k-2}(q t/2) +2I_{k}(q t/2) + I_{k+2}(q t/2)\right]I_{k + 2l}(q t'/2).
\nonumber
\end{align}
Additionally, the modified Bessel functions can be expanded in a Chebyshev series as \cite{W1962}
\begin{align}
I_{\nu}(ax) &= \sum_{n=0}^{\infty} C_n(\nu,a)T_n(x), \nonumber\\
C_n(\nu, a) &= \ii^{n-\nu}\epsilon_n \varphi_n(\nu)I_{\frac{n+\nu}{2}}\left(\frac{a}{2}\right)I_{\frac{n-\nu}{2}}\left(\frac{a}{2}\right),
\nonumber
\end{align}
with $\epsilon_n=2-\delta_{n0}$ and
\begin{align}
\varphi_n(\nu) &= \begin{cases}
(-1)^{n/2}\cos{(\pi\nu/2)}, & \text{for even $n$},\\
(-1)^{(n-1)/2}\sin{(\pi\nu/2)}, & \text{for odd $n$}.
\end{cases}
\nonumber
\end{align}
Using these results, we can readily expand the even-order modified Bessel functions and its derivatives in terms of Chebyshev polynomials $T_m$ as
\begin{align}
I_{2l}(q\abs{t-t'}/2) &= \sum_{m = 0}^{\infty}D^{(0)}_{m}(l,q, t)T_m(t'), \nonumber\\
\dfrac{\partial}{\partial t}I_{2l}(q\abs{t-t'}/2) &= \sum_{m = 0}^{\infty}D^{(1)}_{m}(l,q, t)T_m(t'),\nonumber\\
\dfrac{\partial^2}{\partial t^2}I_{2l}(q\abs{t-t'}/2) &= \sum_{m = 0}^{\infty}D^{(2)}_{m}(l,q, t)T_m(t'),
\nonumber
\end{align}
where the expansion coefficients are defined as
\begin{align}
D^{(0)}_{m}(l,q, t)&=\sum_{k = -\infty}^{\infty}(-1)^k C_{m}(k+2l,q) I_{k}(q t/2), \nonumber\\
D^{(1)}_{m}(l,q, t)&=-\dfrac{q}{4}\sum_{k = -\infty}^{\infty}(-1)^k \left[C_m(k+2l+1,q)+C_m(k+2l-1,q)\right] I_{k}(q t/2), \nonumber\\
D^{(2)}_{m}(l,q, t)&=\dfrac{q^2}{16}\sum_{k = -\infty}^{\infty}(-1)^k \left[C_m(k+2l+2,q)+2C_m(k+2l,q)+C_m(k+2l-2,q)\right] I_{k}(q t/2).
\nonumber
\end{align}
Finally, we can obtain the Chebyshev expansion of the functions in the kernel of the integral equation in Eq.\ \eqref{eq:mateq1} as
\begin{align}
K_{0}(q\abs{t-t'}/2) &=\sum_{m = 0}^{\infty}\left[-\ln{\abs{t-t'}}D^{(0)}_{m}(0,q, t) +K_m^{(0)}(q,t)\right]T_m(t'), \nonumber\\
\dfrac{\partial}{\partial t}K_{0}(q\abs{t-t'}/2) &=\sum_{m = 0}^{\infty}\left[-\dfrac{1}{t-t'}D^{(0)}_{m}(0,q, t) - \ln{\abs{t-t'}}D^{(1)}_{m}(0,q, t)+K_m^{(1)}(q,t)\right]T_m(t'), \nonumber\\
\dfrac{\partial^2}{\partial t^2}K_{0}(q\abs{t-t'}/2) &=\sum_{m = 0}^{\infty}\left[\dfrac{1}{(t-t')^2}D^{(0)}_{m}(0,q, t) - \dfrac{2}{t-t'}D^{(1)}_{m}(0,q, t)-\ln{\abs{t-t'}}D^{(2)}_{m}(0,q, t) +K_m^{(2)}(q,t)\right]T_m(t'),
\nonumber
\end{align}
where we have defined the quantities
\begin{align}
K_m^{(n)}(q,t) &= \sum_{k=0}^{\infty}\zeta_{k}(q)D^{(n)}_{m}(k,q, t),\quad\quad\text{for $n=0,1,2$,} \nonumber\\
\zeta_{k}(q) &= \begin{cases}
-\left[\ln{(q/4)} + \gamma\right], & k = 0, \nonumber\\
2/k, & \text{otherwise}.
\end{cases}
\end{align}
It is also convenient to expand the solutions for $\Eeps_{sx}$ and ${\Eeps}_{sy}$ in terms of the Chebyshev polynomials $U_{n}(t)$ as
\begin{align}
\begin{bmatrix}
\Eeps_{sx}(t)\\
\ii{\Eeps}_{sy}(t)
\end{bmatrix} = \sum_{n = 0}^\infty \sqrt{1-t^2}\;U_{n}(t)
\begin{bmatrix}
u_{s,n}\\
v_{s,n}
\end{bmatrix},
\nonumber
\end{align}
which allows us to rewrite Eq.~\eqref{eq:mateq1} in the form
\begin{align}
\dfrac{1}{\eta_{s}}\sum_{n = 0}^\infty \sqrt{1-t^2}U_{n}(t)\begin{bmatrix}
u_{s,n}\\
v_{s,n}
\end{bmatrix}
 =
 \sum_{p = 0}^\infty\int_{-1}^{1} \begin{bmatrix}
 4\partial_{t}^2 & 2q\partial_{t}\\
 -2q\partial_{t} & -q^2
 \end{bmatrix}
K_{0}\left(q\abs{t-t'}/2\right)\sqrt{1-t^{\prime 2}}U_{p}(t')\begin{bmatrix}
u_{s,p}\\
v_{s,p}
\end{bmatrix} dt'.
\nonumber
\end{align}
Using the identities 
\begin{align}
R_{mp} &= \int_{-1}^{1}\dfrac{T_{m}(t')T_{p}(t')}{\sqrt{1-t^{\prime 2}}}dt'  = \begin{cases}
\pi, & m = p = 0, \\
\dfrac{\pi}{2}\delta_{mp}, & m\neq0, p\neq0,
\end{cases} \nonumber\\
L_{m}(t) &= \int_{-1}^{1}\ln{\abs{t-t'}}\dfrac{T_{m}(t')}{\sqrt{1-t^{\prime 2}}}dt' = -\pi T_{m}(t)\begin{cases}
\ln(2), & m = 0,\\
1/m, & m \neq 0,
\end{cases} \nonumber\\
S_{m} &= {\rm sign}\{m\},
\nonumber
\end{align}
as well as the integration properties of the Chebyshev polynomials, after some algebra, we find the results
\begin{align}
F^{(1)}_{mp}(t) =& \int_{-1}^{1}\sqrt{1-t^{\prime 2}}\dfrac{T_m(t')U_p(t')}{(t-t')^2} = -\dfrac{\pi}{2}\left[(p+m+1)U_{p+m}(t) + (p-m+1)U_{\abs{p-m+1}-1}(t)\right], \nonumber\\
F^{(2)}_{mp}(t) =& \int_{-1}^{1}\sqrt{1-t^{\prime 2}}\dfrac{T_m(t')U_p(t')}{(t-t')} = \dfrac{\pi}{2}\left[T_{p+m+1}(t) + S_{p-m+1}T_{\abs{p-m+1}}(t)\right], \nonumber\\
F^{(3)}_{mp}(t) =& \int_{-1}^{1} \ln{\abs{t-t'}}\sqrt{1-t^{\prime 2}}T_m(t')U_p(t') dt'=\dfrac{1}{4}\bigg[L_{p+m}(t)+L_{\abs{m-p}}(t) - L_{m+p+2}(t) - L_{\abs{m-p-2}}(t)\bigg], \nonumber\\
F^{(4)}_{mp} =& \int_{-1}^{1}\sqrt{1-t^{\prime 2}}T_m(t')U_p(t') dt' = \dfrac{1}{2}\left[R_{m,p} - R_{m, p+2}\right]. \nonumber
\end{align}
With this notation, the integral eigenvalue problem reduces to
\begin{equation}
\dfrac{1}{\eta_{s}}\sum_{n= 0}^\infty U_{n}(t) \sqrt{1-t^2}\begin{bmatrix}
u_{s,n}\\
v_{s,n}
\end{bmatrix}
 =\sum_{p = 0}^\infty\begin{bmatrix}
 M^{11}_{p}(q,t) & M^{12}_{p}(q,t)\\
 M^{21}_{p}(q,t) & M^{22}_{p}(q,t)
\end{bmatrix}
\begin{bmatrix}
u_{s,p}\\
v_{s,p}
\end{bmatrix},
\nonumber
\end{equation}
where we introduced the definitions
\begin{align}
M^{11}_{p}(q,t) &= 4\sum_{m=0}^{\infty}\left[ D_{m}^{(0)}(0,q, t)F_{mp}^{(1)}(t) - 2D_{m}^{(1)}(0,q, t)F_{mp}^{(2)}(t) - D_{m}^{(2)}(0,q, t)F_{mp}^{(3)}(t) + K_{m}^{(2)}(q, t)F_{mp}^{(4)}\right], \nonumber\\
M^{12}_{p}(q,t) &= 2q\sum_{m=0}^{\infty}\left[ -D_{m}^{(0)}(0,q, t)F_{mp}^{(2)}(t) - D_{m}^{(1)}(0,q , t)F_{mp}^{(3)}(t)+ K_{m}^{(1)}(q , t)F_{mp}^{(4)}\right], \nonumber\\
M^{21}_{p}(q ,t) &= -M^{12}_{p}(q ,t), \nonumber\\
M^{22}_{p}(q,t) &=  -q^2\sum_{m=0}^{\infty}\left[-D_{m}^{(0)}(0,q ,t)F_{mp}^{(3)}(t)+ K_{m}^{(0)}(q , t)F_{mp}^{(4)}\right]. \nonumber
\end{align}
This eigenvalue problem can be recast as a generalized matrix eigenvalue problem if we choose a set of $N$ collocation points $t_{l} = \cos\left[\pi(l+1)/(N+1)\right]$ with $l = 0,\cdots,(N-1)$. After doing so, we can write 
\begin{equation}
 \dfrac{1}{\eta_{s}} 
 \begin{bmatrix}
\mathcal{A} & \mathbf{0} \\
\mathbf{0} & \mathcal{A}
\end{bmatrix} \cdot 
\begin{bmatrix}
    \mathbf{u}_{s} \\ \mathbf{v}_{s}
\end{bmatrix}
= 
\begin{bmatrix}
 \mathcal{M}^{11} & \mathcal{M}^{12}\\
 \mathcal{M}^{21} & \mathcal{M}^{22}
 \end{bmatrix}
 \cdot
\begin{bmatrix}
    \mathbf{u}_{s} \\ \mathbf{v}_{s}
\end{bmatrix},
\label{eq:matrixeq}
\end{equation}
where $\mathbf{u}_{s}=\begin{bmatrix} u_{s,0} & \cdots & u_{s,N-1} \end{bmatrix}^{\rm T}$, $\mathbf{v}_{s}=\begin{bmatrix} v_{s,0} & \cdots & v_{s,N-1} \end{bmatrix}^{\rm T}$ (the superscript T indicates the transpose),
\begin{equation}
    \mathcal{A} = \begin{bmatrix}
    U_{0}(t_{0})\sqrt{1 - t_{0}^2} & \cdots & U_{N-1}(t_{0})\sqrt{1 - t_{0}^2} \\
    \vdots & & \vdots \\
    U_{0}(t_{N-1})\sqrt{1 - t_{N-1}^2} & \cdots & U_{N-1}(t_{N-1})\sqrt{1 - t_{N-1}^2}
    \end{bmatrix},
\nonumber
\end{equation} 
\begin{equation}
    \mathcal{M}^{\alpha\beta} = \begin{bmatrix}
    M^{\alpha\beta}_{0}(q ,t_{0})
 & \cdots & M^{\alpha\beta}_{N-1}(q ,t_{0}) \\
    \vdots & & \vdots \\
    M^{\alpha\beta}_{0}(q ,t_{N-1}) & \cdots & M^{\alpha\beta}_{N-1}(q ,t_{N-1})
    \end{bmatrix}
\nonumber
\end{equation} 
for $\alpha\beta\in\{11,12,21,22\}$, and $\mathbf{0}$ is a $N\times N$ zero matrix.

The eigenvalues $\eta_{s}$ and eigenvectors $\uv_{s}$ and $\vv_{s}$ can be readily found from Eq.\ \eqref{eq:matrixeq} using standard numerical algebra methods. In general, these methods yield orthonormal eigenvectors with elements $\tilde{u}_{s}$ and $\tilde{v}_{s}$ (we add the tilde here to clarify that these are the orthonormal eigenvectors that come directly from the eigenvalue equation) that obey the property
\begin{equation}
\sum_{n=0}^{N-1} \left( \tilde{u}_{s,n}^* \tilde{u}_{s',n}  + \tilde{v}_{s,n}^* \tilde{v}_{s',n} \right) = \delta_{jj'},
\nonumber
\end{equation}
where $s=\{j,q\}$ and $s'=\{j',q\}$ (i.e., we are dealing with a fixed value of $q$). However, we note that it is convenient to normalize the obtained eigenvectors in a way that ensures the orthonormality conditions of the fields $\Eeps_{sx}$ and $\Eeps_{sy}$. This can be done by dividing the eigenvectors $\tilde{u}_{s}$ and $\tilde{v}_{s}$ by a factor $\sqrt{A_s}$ with
\begin{align}
    A_{s} &= \sum_{m = 0}^{N-1}\sum_{n = 0}^{N-1} U_{mn} \left( \tilde{u}_{s,m}^* \tilde{u}_{s,n} + \tilde{v}_{s,m}^* \tilde{v}_{s,n} \right), \nonumber\\
    U_{mn} &= 
    \begin{cases}
    0, & |m-n|=1, \\
    \dfrac{(1+m)(1+n)(1+(-1)^{m+n})}{(1-m+n)(1+m-n)(1+m+n)(3+m+n)}, & {\rm otherwise}.
    \end{cases}
\nonumber
\end{align}
In this way, the orthonormality of the fields is ensured, so we have
\begin{equation}
\int_{-1/2}^{1/2} d\theta_x \left[ \Eepsv^*_{s}(\theta_x) \cdot \Eepsv_{s'}(\theta_x) \right] = \sum_{m = 0}^{N-1}\sum_{n = 0}^{N-1} U_{mn} \left( {u}_{s,m}^* {u}_{s',n} + {v}_{s,m}^* {v}_{s',n} \right) = \delta_{jj'},
\nonumber
\end{equation} 
where $u_{s,n}=\tilde{u}_{s,n}/\sqrt{A_{s}}$ and $v_{s,n}=\tilde{v}_{s,n}/\sqrt{A_{s}}$.

After the expansion coefficients are found, the fields and related physical quantities can be computed analytically. We present a set of numerically obtained eigenvalues and eigenvectors in Tables \ref{tab:eta}-\ref{tab:uvj:q0} below. We also show in supplementary Figure\ \ref{fig:S5} the $q$-dependence of the first six modes of $\eta_{s}$, as well as the spatial profile of the charge distribution and the associated electric fields for the first three modes and different values of $q$. Once the eigenvectors are normalized, we can also obtain the total charge density of the $j^{\rm th}$ mode in the ribbon as $\rho_s (\thetav)=\rho_s (\theta_x) \ee^{\ii q \theta_y}$, where the $q$-dependent $\rho_{s}(\theta_x)$ function is given by
\begin{equation}
\rho_{s} (\theta_x) = \sum_{n = 0}^{N-1}\bigg[-2u_{s,n}(n+1)\dfrac{T_{n+1}(2\theta_x)}{\sqrt{1-4\theta_x^2}} + q v_{s,n}\sqrt{1-4\theta_x^2}U_{n}(2\theta_x)\bigg].
\nonumber
\end{equation}
The Fourier transform of $\rho_{s} (\thetav)$, as a function of $\kparb=k_x \hat{\bf x} + k_y \hat{\bf y}$, can finally be computed analytically, yielding
\begin{equation}
\tilde{\tilde{\rho_j}} (\kparb) = \dfrac{\pi^2}{2}\sum_{n=0}^{\infty}(-\ii)^{n-1}\left\{4(n+1)u_{j,k_yD,n}J_{n+1}\left(\dfrac{k_x D}{2}\right)-\ii (k_y D) v_{j,k_yD,n}\left[J_{n}\left(\dfrac{k_x D}{2}\right) +J_{n+2}\left(\dfrac{k_x D}{2}\right)\right]\right\}.
\nonumber
\end{equation}
Given a certain surface conductivity of the ribbons, $\sigma(\omega)$, we can use the obtained eigenvalues to calculate the dispersion relation of the plasmonic modes by numerically solving the equation \cite{paper181}
\begin{equation}
-\eta_{s} = \chi\frac{{\rm Im}\{ \sigma(\omega) \}}{\bar{\epsilon}\,\omega D},  \nonumber
\end{equation}
where $\bar{\epsilon}=(\epsilon_{\rm top}+\epsilon_{\rm bot})/2$ is the average permittivity of the materials above and below the ribbon, and we neglect inelastic losses. Considering for simplicity the Drude conductivity
\begin{equation}
    \sigma(\omega) = \frac{e^2}{\pi \hbar^2} \frac{\ii \EF}{\omega+\ii \gamma},  \nonumber
\end{equation}
defined in terms of the Fermi energy $\EF$ and a phenomenological inelastic decay rate $\gamma\ll\omega$, the dispersion relation admits the solution
\begin{equation}
\label{eq:grapheneDrudeDR}
    \omega_{\rm p}^{(s)} = \frac{e}{\hbar} \frac{1}{\sqrt{\pi(-\eta_{s})}} \sqrt{\frac{\EF}{D}}.
\end{equation}
We represent the resulting dispersion relations of the first six plasmon modes of the graphene ribbon in supplementary Figure\ \ref{fig:S5}.

\section{Multiple plasmon exchanges in extended films}
\label{secS2}

Two-plasmon exchanges can be approximately described through the relation
\begin{equation}
\Gamma_{\rm EELS}^{(2)}(\kvp, \w) \approx \Gamma_{\rm EELS}^{(1)}(\kvp, \w) + \dfrac{1}{2}\int d^2\kvp'\int d\w'\Gamma_{\rm EELS}^{(1)}(\kvp - \kvp', \w - \w')\Gamma_{\rm EELS}^{(1)}(\kvp', \w'),
\label{eq:conv}
\end{equation}
where
\begin{equation}
\Gamma_{\rm EELS}^{(1)}(\kpar,\omega) = \dfrac{2e^2}{\hbar \pi^2 v^2} \dfrac{k_{\parallel}}{(k_{\parallel}^2 + \omega^2/v^2)^2} {\rm Im}\{r_p(\kpar, \omega)\}
\;[n_\Te(\omega)+1]
\label{eq:P1}
\end{equation}
is the single-plasmon interaction probability presented in the Methods section and involving the Bose-Einstein distribution function $n_\Te(\omega)$ at the electronic temperature $\Te$. The integral in Eq.\ (\ref{eq:conv}) is computationally demanding, so we simplify it by using the plasmon-pole approximation to the reflection coefficient \cite{paper331}
\begin{equation}
r_p(\kpar,\omega) \approx \frac{\Rp(\omega) \kp(\omega)}{\kpar-\kp(\omega)},
\nonumber
\end{equation}
where $k_p(\omega)$ corresponds to the dispersion relation of the plasmons supported by the graphene sheet and $\Rp(\omega)$ is a dimensionless residue. Disregarding plasmonic losses (i.e, we take the imaginary part of $k_p$ to be infinitesimal), we obtain
 \begin{equation}
 {\rm Im}\{r_p(\kpar, \omega)\} \approx -\ii \pi \Rp(\omega) \kp(\omega) \delta[\kpar-\kp(\omega)].
 \label{eq:Imr}
\end{equation}
Plugging Eqs.\ \eqref{eq:P1} and \eqref{eq:Imr} into Eq.\ \eqref{eq:conv}, we find
\begin{align}
\Gamma_{\rm EELS}^{(2)}(\kvp, \w) = &\dfrac{-2e^4}{\hbar^2 \pi^2 v^4}\int d\w' \int \kpar' d\kpar' \int d\varphi\;\frac{|\kparb-\kparb'| \Rp(\w-\w')\kp(\w-\w')}{\left( |\kparb-\kparb'|^2+(\w-\w')^2/v^2 \right)^2}\;\frac{\kpar' \Rp(\w')\kp(\w')}{\left( \kpar'^2+\w'^2/v^2 \right)^2} \label{eq:conv2} \\
\times&[n_\Te(\omega-\omega')+1]\;[n_\Te(\omega)+1]\;\delta[|\kparb-\kparb'|-\kp(\omega-\omega')]\;\delta[\kpar'-\kp(\omega')].
\nonumber
\end{align}
The last delta function in the second line of Eq.\ \eqref{eq:conv2} readily simplifies the $\kpar'$ integral, effectively allowing us to set $\kpar'=\kp(\omega)$. In addition, noticing that $|\kpar-\kp(\omega')|=\sqrt{\kpar^2 + \kp^2(\omega') -2\kpar\kp(\omega')\cos(\varphi)}$ and making the change of variables $u=\cos(\varphi)$, we can rewrite the remaining delta function as
\begin{align}
\delta[|\kparb-\kparb'|-\kp(\omega-\omega')] &= \frac{\kp(\w-\w')}{\kpar \kp(\w')}\delta(u-u_0),
\nonumber
\end{align}
with $u_0 = [\kpar^2 + \kp^2(\w')-\kp^2(\w-\w')]/[2\kpar \kp(\w')]$. After some straightforward algebra, we find that the two-plasmon loss probability reduces to
\begin{align}
\Gamma_{\rm EELS}^{(2)}(\kvp, \w) = \frac{4 e^4}{\pi^2 \hbar^4 v^4} \int  \frac{ d\w'\; \Pi(\w-\w')\;\Pi(\w')}{\sqrt{4\kpar^2 \kp^2(\w') - (\kpar^2 + \kp^2(\w') - \kp^2(\w-\w'))^2}} \Theta\left[1-\left|\frac{\kpar^2 + \kp^2(\w') - \kp^2(\w-\w')}{2\kpar \kp(\w')}\right|\right],
\nonumber
\end{align}
where
\begin{equation}
\Pi(\w) = \frac{\kp(\w)^3 \Rp(\w)^3}{\kp(\w)^2+\w^2/v^2}\;[n_\Te(\omega)+1],
\nonumber
\end{equation}
with the Heaviside function $\Theta$ originating in the integral over $u$, which is zero unless $|u_0|\leq 1$. We use this expression to obtain supplementary Figure\ \ref{fig:S4}, as it only involves a one-dimensional integral, so it is fast to compute.

\begin{table*}[htbp]
\caption{Fitting parameters of the first six ribbon eigenvalues $1/\eta_{j,q} = \sum_m a_m q^m$ as a function of $q$. Parameters with an absolute value smaller that $10^{-4}$ are omitted.}
\label{tab:eta}
\footnotesize
\renewcommand{\arraystretch}{1}
\begin{tabular}{|C{1cm}|C{\cww}|C{\cww}|C{\cww}|C{\cww}|C{\cww}|C{\cww}|C{\cww}|C{\cww}|C{\cww}|}
\hline
$j$ & $a_{-4}$ & $a_{-3}$ & $a_{-2}$ & $a_{-1}$ & $a_{0}$ & $a_{1}$ & $a_{2}$ & $a_{3}$ & $a_{4}$ \\
\hline
$1$ & $1.14\times 10^{-4}$ & $-1.5\times 10^{-2}$ & $3.82\times 10^{-1}$ & $-2.09$ & $4.63$ & $-6.13$ & $7.71\times 10^{-2}$ & $-2.66\times 10^{-3}$ &  --  \\ \hline   $2$ & $1.46\times 10^{-2}$ & $-1.55$ & $9.19$ & $-1.82\times 10^1$ & $3.72\times 10^{-1}$ & $-4.65$ & $-5.96\times 10^{-2}$ & $2.45\times 10^{-3}$ &  --  \\ \hline   $3$ & $1.25\times 10^{-2}$ & $-1.32$ & $7.39$ & $-1.29\times 10^1$ & $-2.63\times 10^1$ & $-1.81$ & $-3.16\times 10^{-1}$ & $1.01\times 10^{-2}$ & $-1.21\times 10^{-4}$ \\ \hline   $4$ & $3.18\times 10^{-3}$ & $-3.36\times 10^{-1}$ & $1.82$ & $-2.95$ & $-5.27\times 10^1$ & $-1.18\times 10^{-1}$ & $-3.46\times 10^{-1}$ & $9.48\times 10^{-3}$ & $-1.02\times 10^{-4}$ \\ \hline   $5$ &  --  & $-8.43\times 10^{-3}$ & $-9.83\times 10^{-3}$ & $2.41\times 10^{-1}$ & $-7.45\times 10^1$ & $3.61\times 10^{-1}$ & $-3.05\times 10^{-1}$ & $7.24\times 10^{-3}$ &  --  \\ \hline   $6$ & $-1.14\times 10^{-3}$ & $1.21\times 10^{-1}$ & $-7.16\times 10^{-1}$ & $1.4$ & $-9.49\times 10^1$ & $4.37\times 10^{-1}$ & $-2.55\times 10^{-1}$ & $5.14\times 10^{-3}$ &  --  \\ \hline
\end{tabular}
\end{table*}

\begin{table*}[htbp]
\caption{Ribbon eigenvectors $u_j$ and $v_{j}$ for $q=0$. For each mode $j=0$ to 10, the first 25 elements $v_{0,n}$ and $u_{j>0,n}$ of the corresponding eigenvector are presented. We omit elements $u_{0,n}$ and $v_{j>0,n} \approx 0$, which take negligible values. Vector elements smaller that $10^{-4}$ are also omitted.}
\label{tab:uvj:q0}
\footnotesize
\renewcommand{\arraystretch}{1}
\begin{tabular}{|C{1cm}|C{\cw}|C{\cw}|C{\cw}|C{\cw}|C{\cw}|C{\cw}|C{\cw}|C{\cw}|C{\cw}|C{\cw}|C{\cw}|C{\cw}|}
\hline
$n$ & $v_{0,n}$ & $u_{1,n}$ & $u_{2,n}$ & $u_{3,n}$ & $u_{4,n}$ & $u_{5,n}$ & $u_{6,n}$ & $u_{7,n}$ & $u_{8,n}$ & $u_{9,n}$ & $u_{10,n}$\\
\hline
$1$ & $-1.2732$ & $-1.2004$ &  --  & $0.3040$ &  --  & $-0.1739$ &  --  & $0.1217$ &  --  & $0.0936$ &  --  \\ \hline   $2$ &  --  &  --  & $-1.2517$ &  --  & $-0.5189$ &  --  & $-0.3226$ &  --  & $-0.2331$ &  --  & $0.1821$ \\ \hline   $3$ & $-0.4237$ & $0.1050$ &  --  & $1.1736$ &  --  & $-0.6517$ &  --  & $0.4413$ &  --  & $0.3311$ &  --  \\ \hline   $4$ &  --  &  --  & $0.3052$ &  --  & $-1.0124$ &  --  & $-0.7145$ &  --  & $-0.5280$ &  --  & $0.4132$ \\ \hline   $5$ & $-0.2535$ & $0.0021$ &  --  & $-0.4989$ &  --  & $-0.7965$ &  --  & $0.7073$ &  --  & $0.5741$ &  --  \\ \hline   $6$ &  --  &  --  & $-0.0149$ &  --  & $0.6716$ &  --  & $-0.5425$ &  --  & $-0.6365$ &  --  & $0.5763$ \\ \hline   $7$ & $-0.1802$ & $0.0010$ &  --  & $0.0616$ &  --  & $0.8005$ &  --  & $0.2769$ &  --  & $0.5133$ &  --  \\ \hline   $8$ &  --  &  --  & $0.0027$ &  --  & $-0.1389$ &  --  & $0.8758$ &  --  & $-0.0184$ &  --  & $0.3510$ \\ \hline   $9$ & $-0.1393$ & $0.0004$ &  --  & $-0.0072$ &  --  & $-0.2396$ &  --  & $-0.8897$ &  --  & $-0.2104$ &  --  \\ \hline   $10$ &  --  &  --  & $0.0008$ &  --  & $0.0193$ &  --  & $-0.3559$ &  --  & $0.8415$ &  --  & $-0.3945$ \\ \hline   $11$ & $-0.1131$ & $0.0002$ &  --  & $-0.0009$ &  --  & $0.0430$ &  --  & $0.4752$ &  --  & $-0.7361$ &  --  \\ \hline   $12$ &  --  &  --  & $0.0004$ &  --  &  --  &  --  & $0.0820$ &  --  & $-0.5860$ &  --  & $-0.5826$ \\ \hline   $13$ & $-0.0948$ &  --  &  --  & $-0.0006$ &  --  & $-0.0032$ &  --  & $-0.1372$ &  --  & $0.6752$ &  --  \\ \hline   $14$ &  --  &  --  & $0.0002$ &  --  & $0.0007$ &  --  & $-0.0103$ &  --  & $0.2084$ &  --  & $0.7328$ \\ \hline   $15$ & $-0.0813$ &  --  &  --  & $-0.0003$ &  --  & $0.0010$ &  --  & $0.0235$ &  --  & $-0.2919$ &  --  \\ \hline   $16$ &  --  &  --  & $0.0001$ &  --  & $0.0004$ &  --  & $0.0019$ &  --  & $-0.0453$ &  --  & $-0.3828$ \\ \hline   $17$ & $-0.0708$ &  --  &  --  & $-0.0002$ &  --  & $0.0004$ &  --  & $-0.0039$ &  --  & $0.0774$ &  --  \\ \hline   $18$ &  --  &  --  &  --  &  --  & $0.0003$ &  --  & $0.0004$ &  --  & $0.0081$ &  --  & $0.1212$ \\ \hline   $19$ & $-0.0624$ &  --  &  --  & $-0.0001$ &  --  & $0.0003$ &  --  & $-0.0002$ &  --  & $-0.0157$ &  --  \\ \hline   $20$ &  --  &  --  &  --  &  --  & $0.0002$ &  --  & $0.0003$ &  --  & $-0.0004$ &  --  & $-0.0283$ \\ \hline   $21$ & $-0.0555$ &  --  &  --  &  --  &  --  & $0.0002$ &  --  & $-0.0003$ &  --  & $0.0017$ &  --  \\ \hline   $22$ &  --  &  --  &  --  &  --  & $0.0001$ &  --  & $0.0002$ &  --  & $0.0004$ &  --  & $0.0043$ \\ \hline   $23$ & $-0.0497$ &  --  &  --  &  --  &  --  & $0.0001$ &  --  & $-0.0002$ &  --  & $-0.0006$ &  --  \\ \hline   $24$ &  --  &  --  &  --  &  --  &  --  &  --  & $0.0002$ &  --  & $0.0002$ &  --  & $-0.0010$ \\ \hline   $25$ & $-0.0448$ &  --  &  --  &  --  &  --  &  --  &  --  & $-0.0002$ &  --  & $-0.0002$ &  --  \\ \hline   $26$ &  --  &  --  &  --  &  --  &  --  &  --  & $0.0001$ &  --  & $0.0002$ &  --  & $-0.0002$ \\ \hline   $27$ & $-0.0405$ &  --  &  --  &  --  &  --  &  --  &  --  & $-0.0001$ &  --  & $-0.0002$ &  --  \\ \hline
\end{tabular}
\end{table*}


\end{widetext}

\section{Acknowledgments}

This work has been supported in part by the European Research Council (Advanced Grant 789104-eNANO), the Spanish MINECO (MAT2017-88492-R and SEV2015-0522), the European Commission (Graphene Flagship 696656), the Catalan CERCA Program, and Fundaci\'o Privada Cellex. F.C. acknowledges support from the ERC consolidator grant ISCQuM. E.J.C.D. acknowledges financial support from “la Caixa” (INPhINIT Fellowship Grant 1000110434, LCF/BQ/DI17/11620057) and the EU (Marie Sk\l{}odowska-Curie Grant 713673).


\section{Supplementary Figures}

\begin{figure*}[htbp]
\centering
\includegraphics[width=0.95\textwidth]{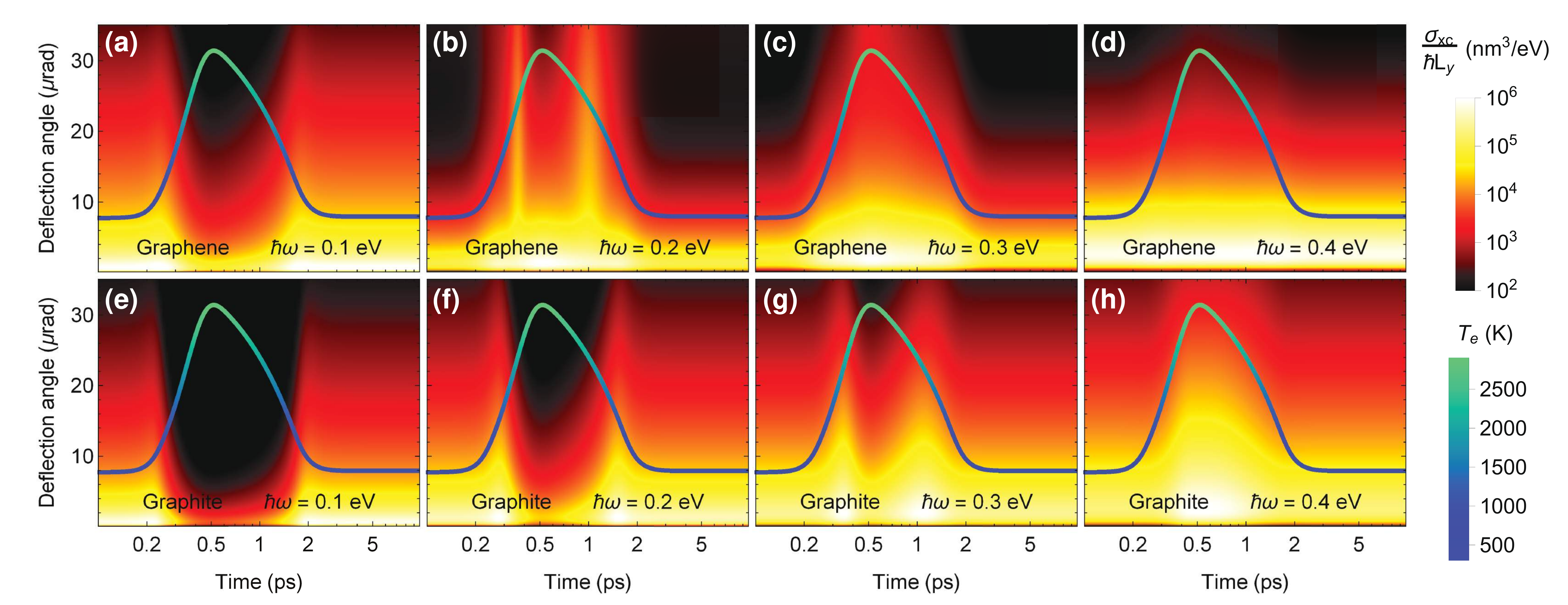}
\caption{Same as Figure\ \ref{Fig1}c, but for (a-d) graphene (0.2\,eV Fermi energy, 4\,meV intrinsic damping) and (e-h) graphite (undoped, 10 monolayers) 100-nm-wide ribbons with different selected energy losses $\hbar\omega$ (see labels), illuminated by a Gaussian pump pulse of 166.5\,fs FWHM delivering $100\,{\rm GW/m^2}$ peak absorption power density, and probed with $100\,$keV electrons. The density plots show the inelastic cross section $\sigma_{\rm xc}(\kparb,\omega)$ resolved in lateral momentum $\hbar\kparb$ and energy $\hbar\omega$ transfers.}
\label{fig:S1}
\end{figure*}

\begin{figure*}[htbp]
\centering
\includegraphics[width=0.95\textwidth]{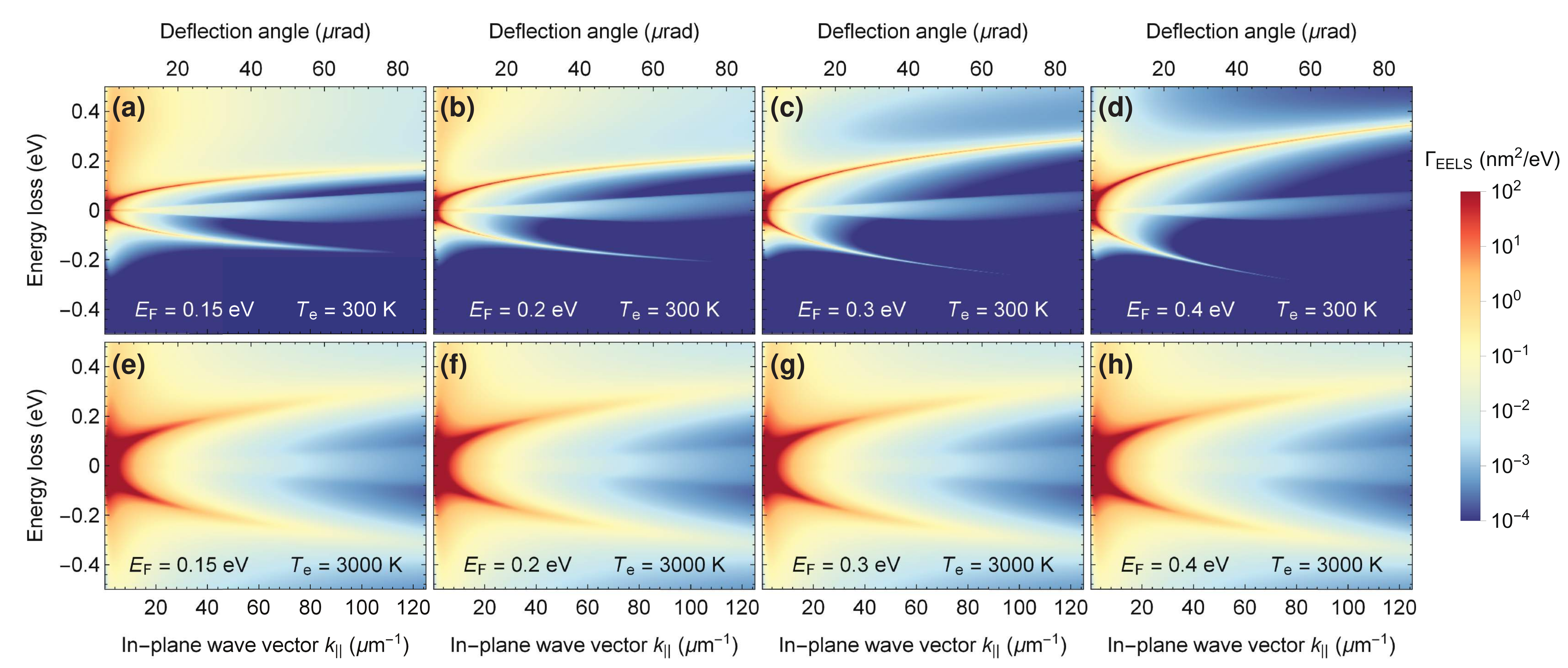}
\caption{Same as Figure\ \ref{Fig2}c,d, but for different Fermi energies (see labels). We consider low and high temperatures in the upper and lower plots, respectively.}
\label{fig:S2}
\end{figure*}

\begin{figure*}[htbp]
\centering
\includegraphics[width=0.95\textwidth]{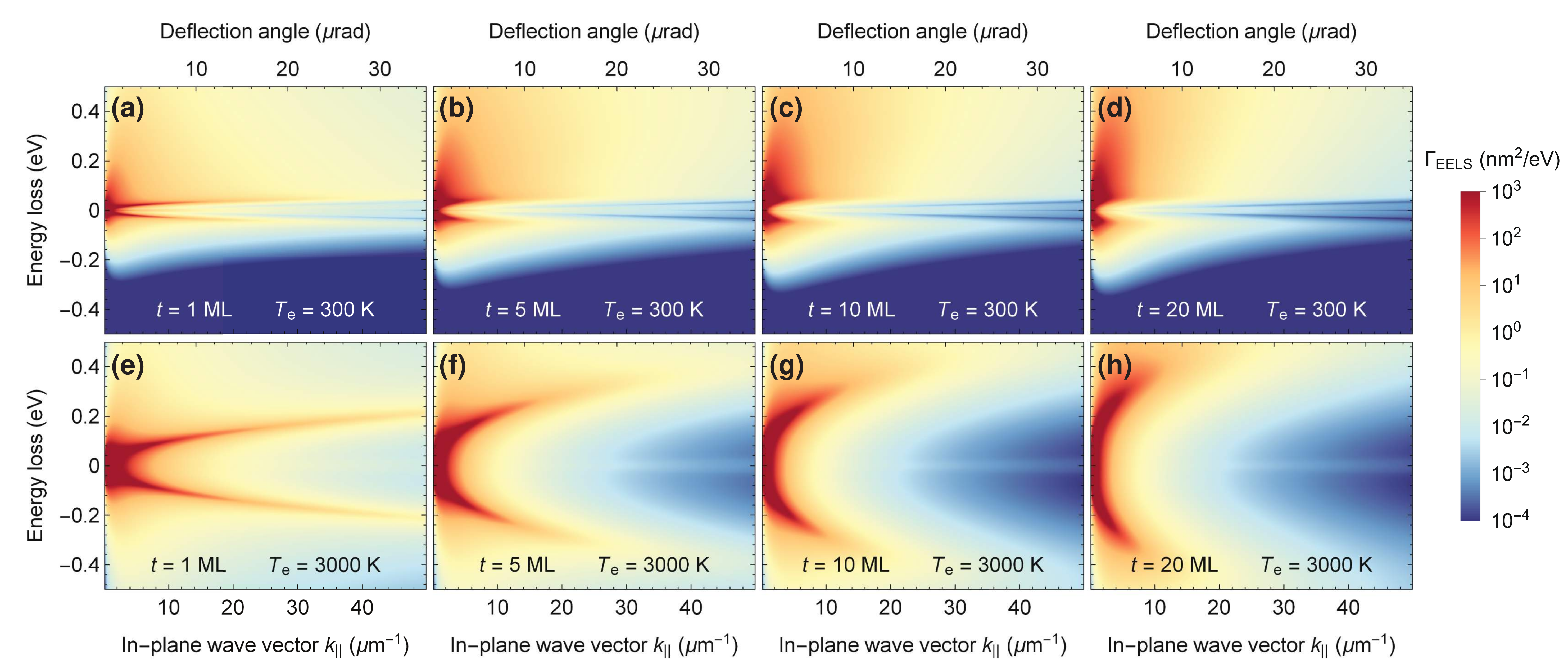}
\caption{Same as Figure\ \ref{Fig2}c,d, but for undoped multilayer graphene films with different numbers of carbon monolayers (MLs, see labels). We consider low and high temperatures in the upper and lower plots, respectively.}
\label{fig:S3}
\end{figure*}

\begin{figure*}[htbp]
\centering
\includegraphics[width=0.86\textwidth]{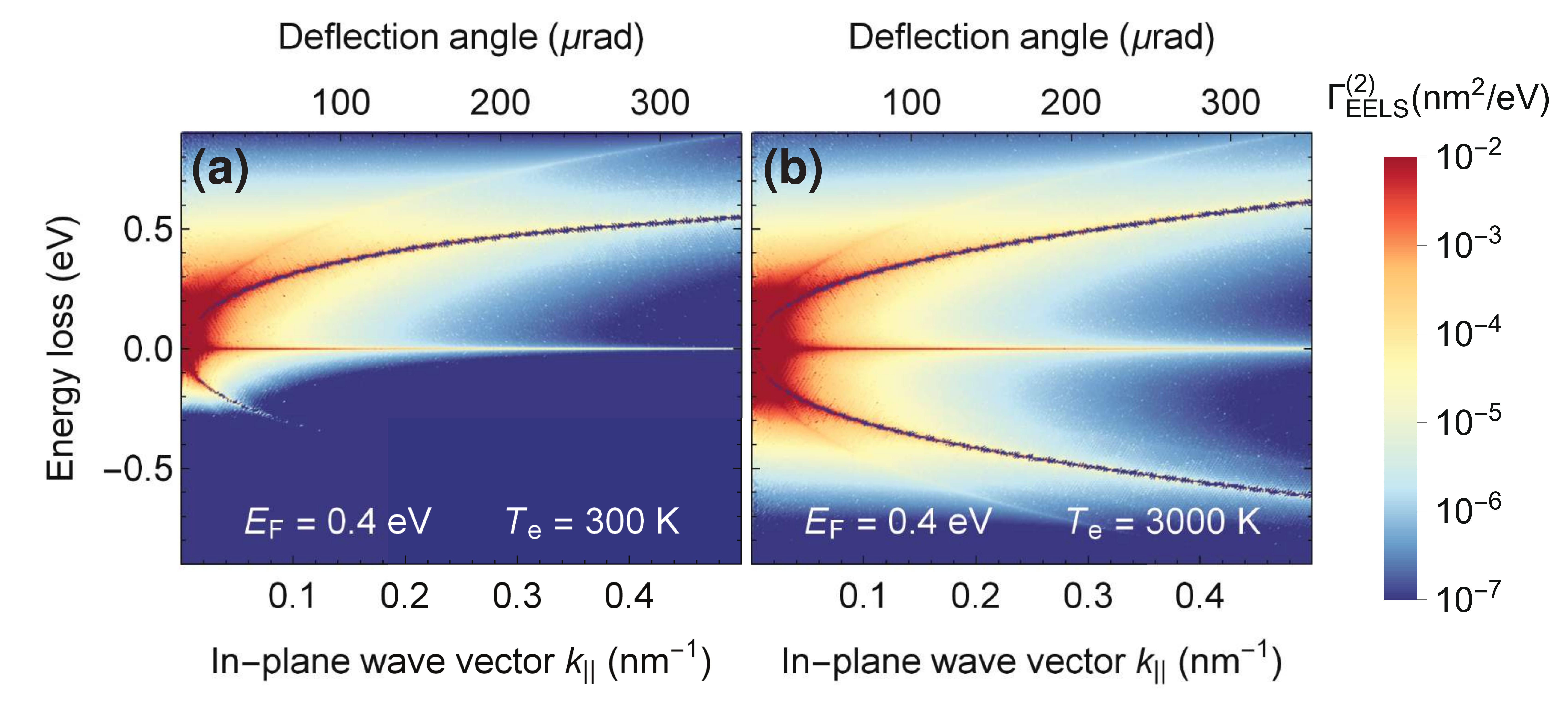}
\caption{Same as Figure\ \ref{Fig2}c,d, but a Fermi energy $\EF=0.4\,$eV and including multiple scattering events (see Sec.\ \ref{secS2} for details of the calculation). The first replica of the plasmon dispersion is clearly discernible. The dominant plasmon feature is now sharper than in Figure\ \ref{Fig2} because we are neglecting the intrinsic graphene damping. A faint plasmon replica indicates the two-plasmon excitation processes.}
\label{fig:S4}
\end{figure*}

\begin{figure*}[htbp]
\centering
\includegraphics[width=\textwidth]{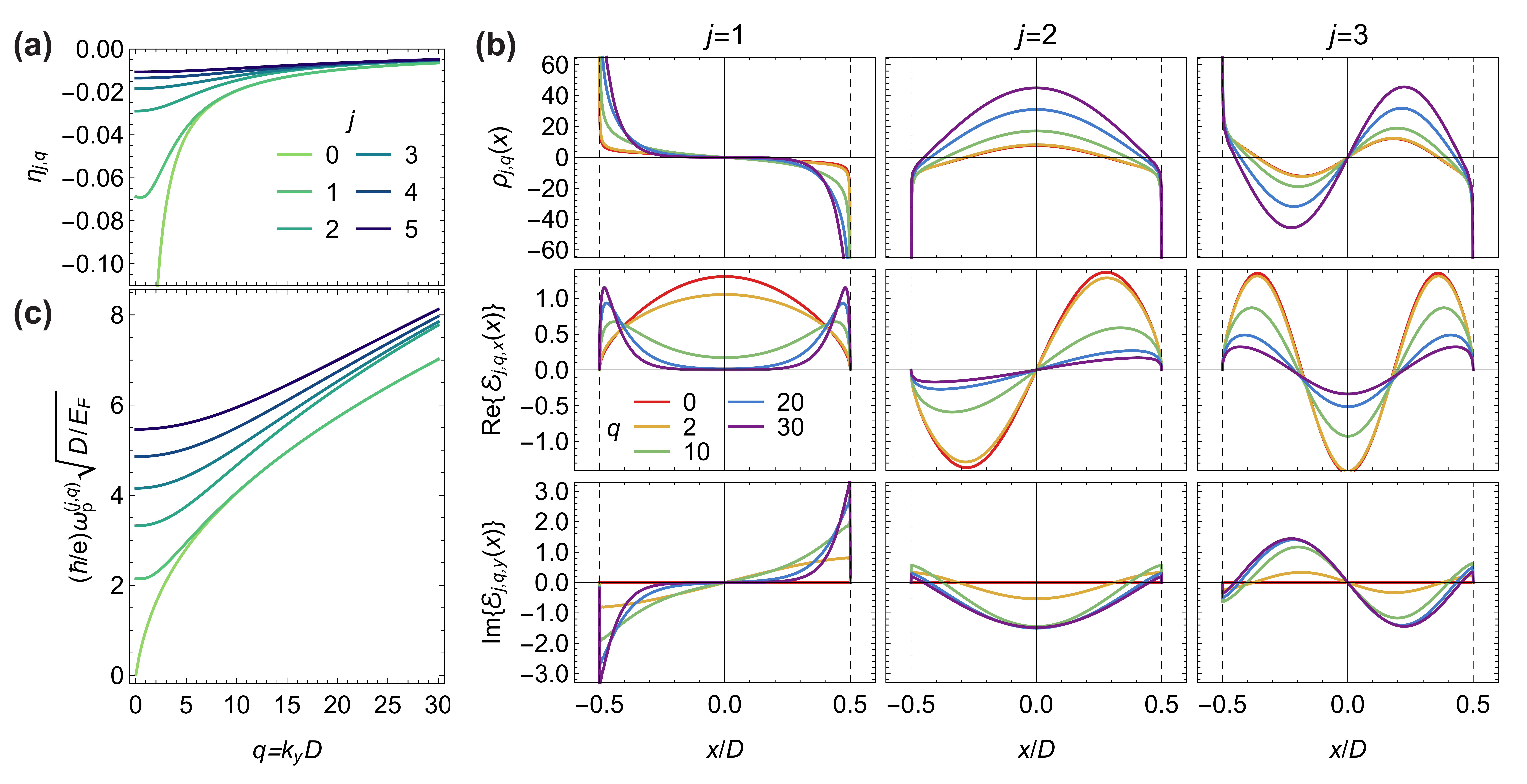}
\caption{Electrostatic egienmodes of a ribbon. (a) Momentum dependence of the eigenvalues $\eta_{s}$ (with $s=\{j,q\}$) for the six lowest-order modes. (b) The corresponding charge distributions (top, $\rho_j(x)$) and electric fields (middle, $\mathcal{E}_{sx}(x)$; bottom, $\mathcal{E}_{sy}(x)$) of modes $j=1$-3 (see top labels) for different values of the normalized parallel wave vector $q=k_yD$ as a function of transverse ribbon position coordinate $x$ normalized to the ribbon width $D$. (c) Dispersion relation of the first six plasmonic modes of a graphene ribbon in the Drude model, as calculated from Eq.~\eqref{eq:grapheneDrudeDR}.}
\label{fig:S5}
\end{figure*}

\begin{figure*}[htbp]
\centering
\includegraphics[width=0.95\textwidth]{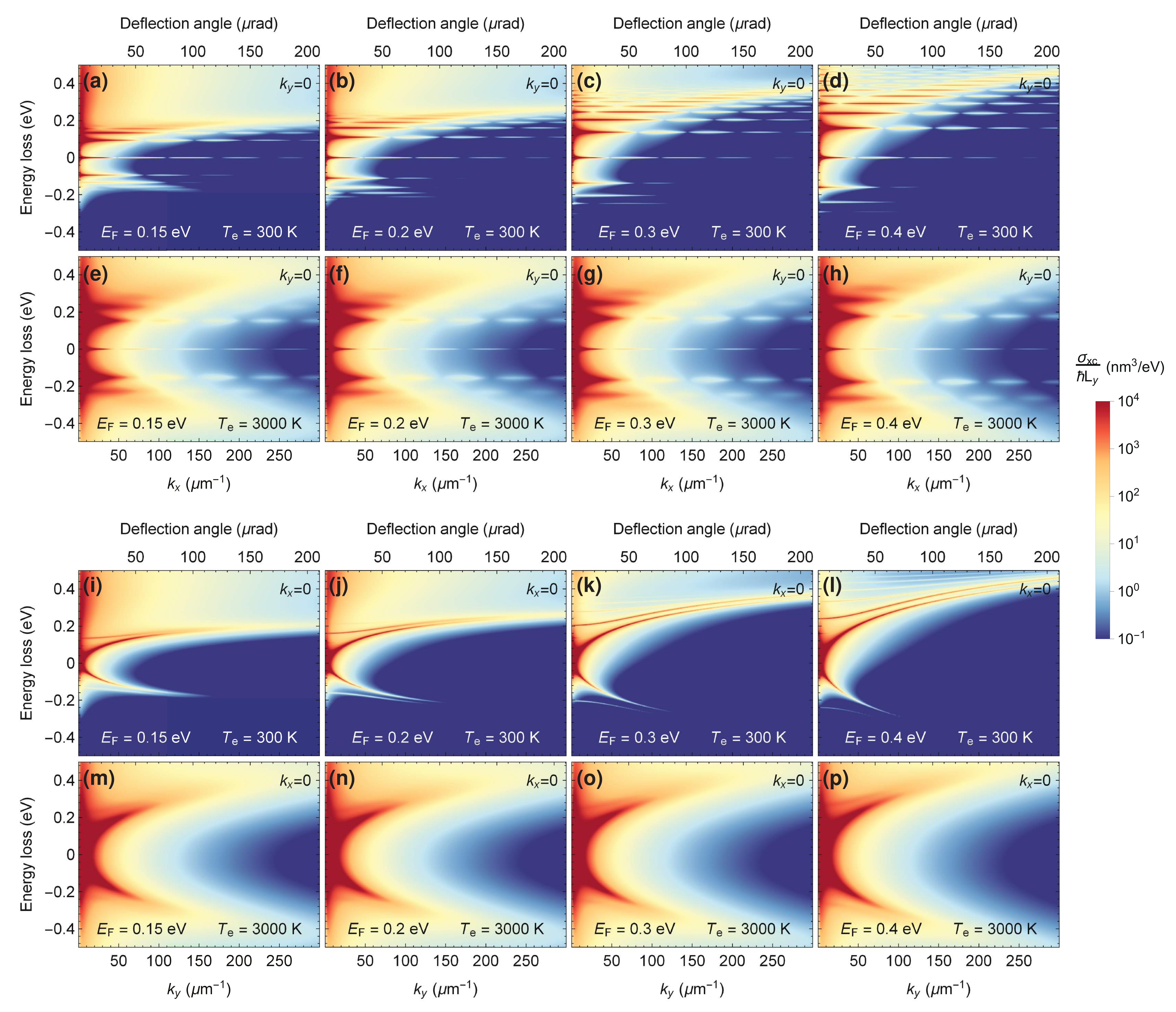}
\caption{Same as Figure\ \ref{Fig3}b-e, but for graphene ribbons with different Fermi energies $\EF$ (see labels). The ribbon width is 100\,nm in all cases.}
\label{fig:S6}
\end{figure*}

\begin{figure*}[htbp]
\centering
\includegraphics[width=0.95\textwidth]{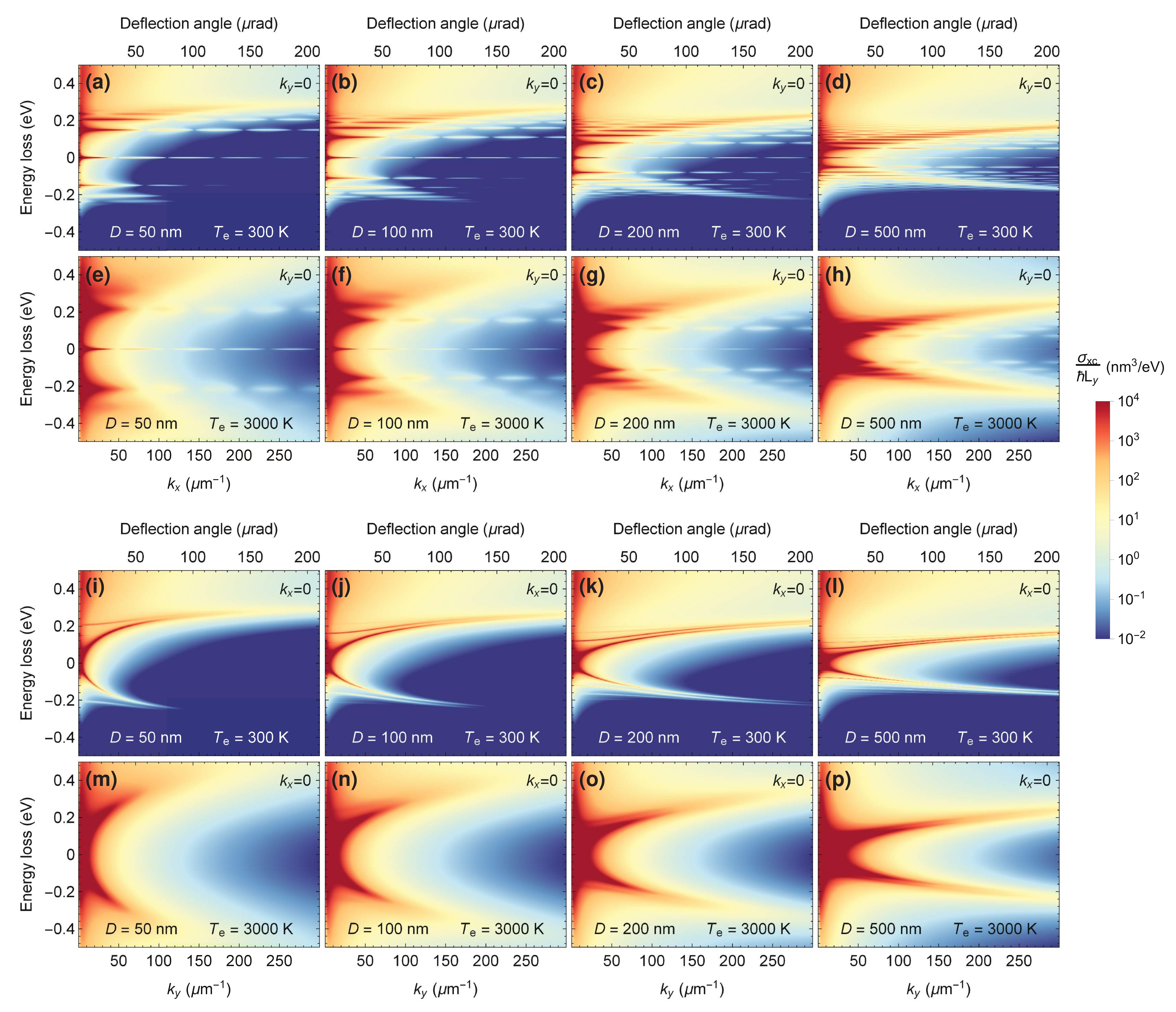}
\caption{Same as Figure\ \ref{Fig3}b-e, but for graphene ribbons with widths $D$ (see labels). The Fermi energy is $\EF=0.4\,$eV in all cases.}
\label{fig:S7}
\end{figure*}

\begin{figure*}[htbp]
\centering
\includegraphics[width=0.95\textwidth]{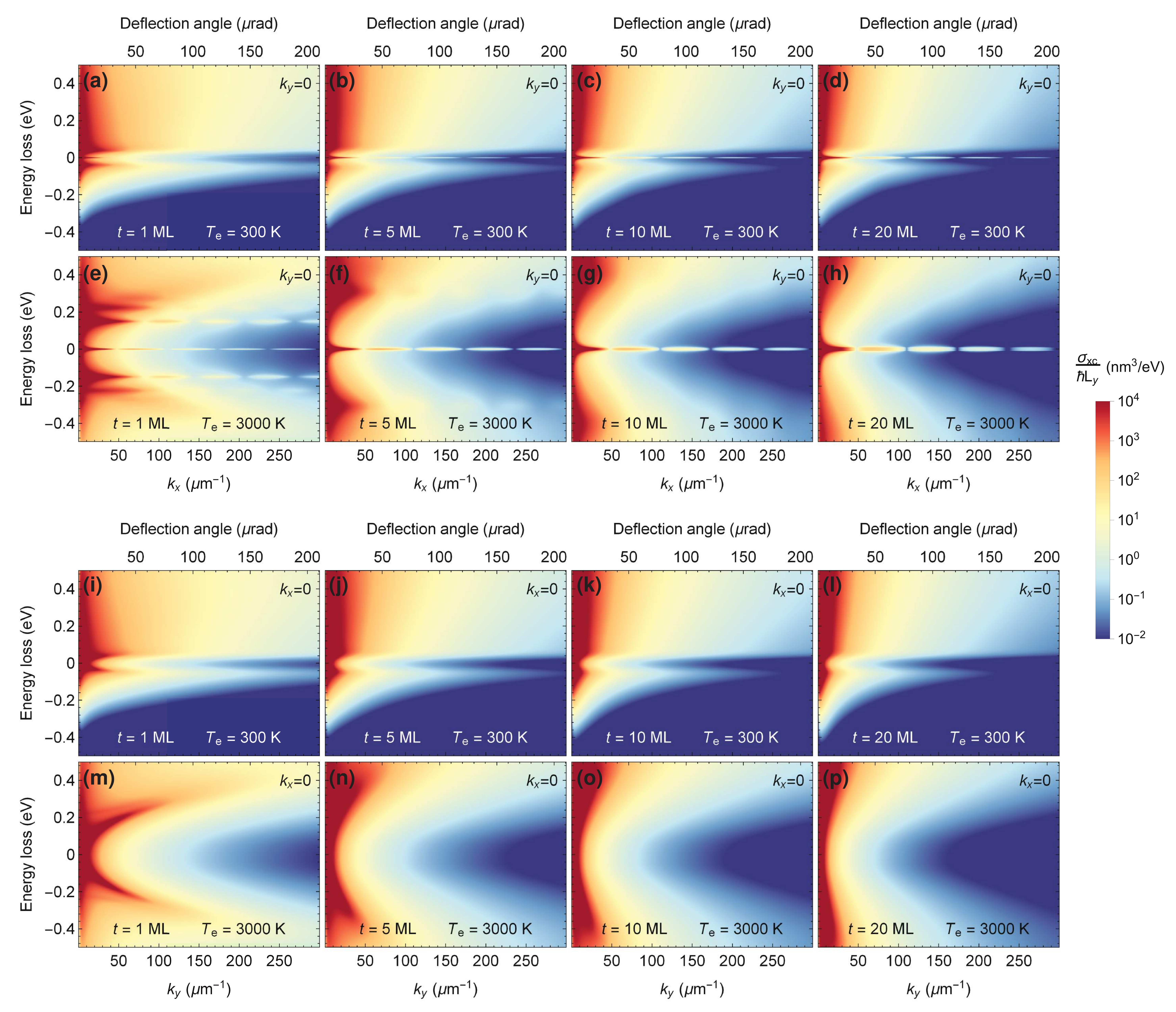}
\caption{Same as Figure\ \ref{Fig3}g-j, but for undoped multilayer graphene ribbons with different numbers of carbon monolayers (MLs, see labels). The ribbon width is 100\,nm in all cases.}
\label{fig:S8}
\end{figure*}

\begin{figure*}[htbp]
\centering
\includegraphics[width=0.9\textwidth]{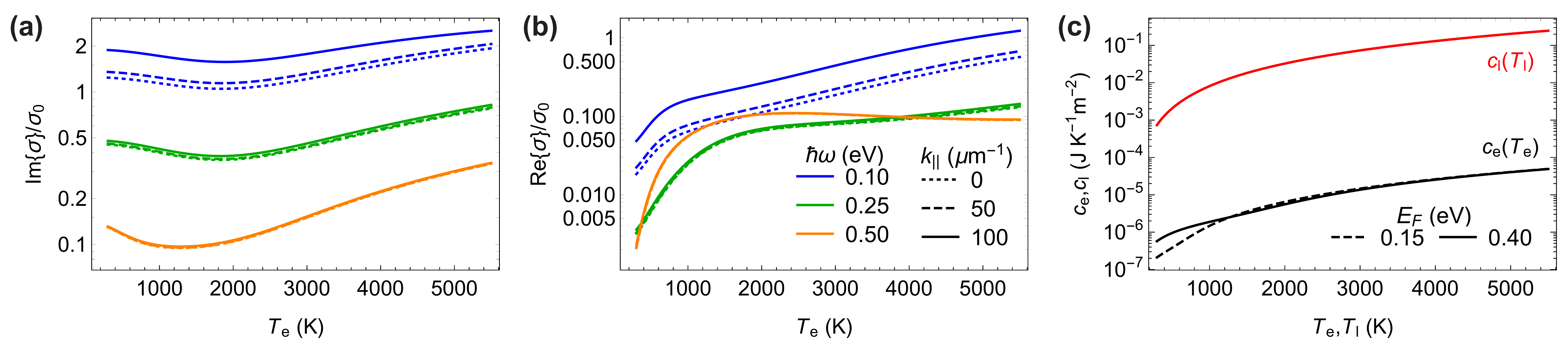}
\caption{Temperature-dependent properties of extended graphene. (a) Imaginary and (b) real parts of the conductivity as a function of electronic temperature $\Te$ for different values of the parallel wave vector $\kpar$ and frequency $\omega$ with a Fermi energy $\EF=0.4\,{\rm eV}$. The conductivity is normalized to $\sigma_0=e^2/\hbar$. (c) Electronic (black) and lattice (red) heat capacities as a function of electronic and lattice temperatures, respectively, for two different values of the Fermi energy (see labels).}
\label{fig:S9}
\end{figure*}

\end{document}